\documentclass[numberedappendix]{emulateapj}
\usepackage{amsmath,amsfonts, amssymb, url, natbib, enumerate, graphicx, mathtools}
\usepackage[normalem]{ulem}
\usepackage{color,url}
\defcitealias{Sharp2010}{S10}
\defcitealias{Pullen2013}{P13}
\begin{document}
\shorttitle{COPSS}
\shortauthors{Keating et al.}
\title{First Results from COPSS: The CO Power Spectrum Survey}
\author{Garrett K. Keating\altaffilmark{1}, Geoffrey C. Bower\altaffilmark{2}, Daniel P. Marrone\altaffilmark{3}, David R. DeBoer\altaffilmark{1}, Carl Heiles\altaffilmark{1}, Tzu-Ching~Chang\altaffilmark{4}, John E. Carlstrom\altaffilmark{5}, Christopher H. Greer\altaffilmark{3}, David Hawkins\altaffilmark{6}, James W. Lamb\altaffilmark{6}, Erik~Leitch\altaffilmark{5,6}, Amber D. Miller\altaffilmark{7,8}, Stephen Muchovej\altaffilmark{6,9}, David P. Woody\altaffilmark{6}}
\altaffiltext{1}{Radio Astronomy Laboratory, 501 Campbell Hall, University of California, Berkeley, CA 94720, USA; karto@astro.berkeley.edu}
\altaffiltext{2}{Academia Sinica Institute of Astronomy and Astrophysics, 645 N. A'ohoku Pl., Hilo, HI 96720, USA; gbower@asiaa.sinica.edu.tw}
\altaffiltext{3}{Steward Observatory, University of Arizona, 933 North Cherry Avenue, Tucson, AZ 85721, USA}
\altaffiltext{4}{Academia Sinica Institute of Astronomy and Astrophysics, P.O. Box 23-141, Taipei 10617, Taiwan}
\altaffiltext{5}{Department of Astronomy \& Astrophysics, Kavli Institute for Cosmological Physics, University 
of Chicago, Chicago, IL 60637, USA}
\altaffiltext{6}{California Institute of Technology, Owens Valley Radio Observatory, Big Pine, CA 93513, USA}
\altaffiltext{7}{Columbia Astrophysics Laboratory, Columbia University, New York, NY 10027, USA}
\altaffiltext{8}{Department of Physics, Columbia University, New York, NY 10027, USA}
\altaffiltext{9}{California Institute of Technology, Department of Astronomy, Pasadena, CA 91125, USA}
\submitted{Accepted for publication in ApJ}
\begin{abstract}
We present constraints on the abundance of carbon-monoxide in the early Universe from the CO Power Spectrum Survey (COPSS). We utilize a data set collected between 2005 and 2008 using the Sunyaev-Zel'dovich Array (SZA), which were previously used to measure arcminute-scale fluctuations of the CMB. This data set features observations of 44 fields, covering an effective area of 1.7 square degrees,  over a frequency range of 27 to 35 GHz. Using the technique of intensity mapping, we are able to probe the CO(1-0) transition, with sensitivity to spatial modes between $k=0.5{-}2\ h\,\textrm{Mpc}^{-1}$ over a range in redshift of $z=2.3{-}3.3$, spanning a comoving volume of $3.6\times10^{6}\ h^{-3}\,\textrm{Mpc}^{3}$. We demonstrate our ability to mitigate foregrounds, and present estimates of the impact of continuum sources on our measurement. We constrain the CO power spectrum to $P_{\textrm{CO}}<2.6\times10^{4}\ \mu\textrm{K}^{2} (h^{-1}\,\textrm{Mpc})^{3}$, or $\Delta^{2}_{\textrm{CO}}(k\! = \! 1 \ h\,\textrm{Mpc}^{-1})<1.3 \times10^{3}\ \mu\textrm{K}^{2}$, at $95\%$ confidence. This limit resides near optimistic predictions for the CO power spectrum. Under the assumption that CO emission is proportional to halo mass during bursts of active star formation, this corresponds to a limit on the ratio of $\textrm{CO}(1{-}0)$ luminosity to host halo mass of $A_{\textrm{CO}}<1.2\times10^{-5}\ L_{\odot}\ M_{\odot}^{-1}$. Further assuming a Milky Way-like conversion factor between CO luminosity and molecular gas mass ($\alpha_{\textrm{CO}}=4.3\ M_{\odot}\ (\textrm{K}\ \textrm{km}\ \textrm{s}^{-1}\ \textrm{pc}^{-2})^{-1}$), we constrain the global density of molecular gas to $\rho_{z\sim3}(M_{\textrm{H}_{2}})\leq 2.8 \times10^{8}\ M_{\odot}\ \textrm{Mpc}^{-3}$.
\end{abstract}
\section{Introduction}\label{sec_intro}
Molecular gas serves a vital role in star formation as the natal material from which stars form. Though the main constituent of this gas is molecular hydrogen, the H$_2$ molecule lacks a permanent dipole moment, making it a poor radiator of energy and hence difficult to observe. Traditionally, the CO molecule -- the next most abundant molecule after H$_2$ -- has been used as a tracer of molecular hydrogen (e.g., \citealt{Wilson1970,Young1982,Young1995,Regan2001,Bolatto2013}). Unlike H$_2$, the CO molecule possesses a permanent dipole moment, with an excitation temperature of $T_{\textrm{ex}} \approx 5.5 \textrm{K}$ for the $J{=}1{\rightarrow}0$ transition, making it ideal for probing the cold, dense gas of molecular clouds. 

Within the local Universe, the CO luminosity ($L_\textrm{CO}$) of galaxies -- and by extension, their molecular gas mass -- shows strong correlation with far-infrared luminosity ($L_{\textrm{FIR}}$), H$\alpha$ emission and Ly$\alpha$ emission; all are strong indicators of star formation within these galaxies \citep{Downes1993,Solomon1997,Kennicutt1998}. This relationship between molecular gas and star formation rates ($SFR$) is typically referred to as the Kennicutt-Schmidt (KS) relationship \citep{Schmidt1959,Kennicutt1998}, and demonstrates a deep connection between the abundance of molecular gas and the formation of stars.

It is unknown how the connection between CO, bulk molecular gas, and star formation evolves over cosmic time. Observations of distant star-forming galaxies suggest that $L_\textrm{CO}$-SFR correlation persists up to $z\lesssim 2$ \citep{Tacconi2013}, implying that the correlation between CO abundance and the amount of molecular gas available for star formation has remained relatively unchanged in the several billion years following peak of cosmic star formation \citep{Hopkins2006}. Prior to this epoch, however, early galaxies (with their short star formation histories) may not contain enough metals to form an appreciable amount of CO, or may possess too little dust to shield the CO from dissociation by UV starlight \citep{Genzel2012,Bolatto2013}. Some models of the conversion factor between CO luminosity and molecular gas mass ($\alpha_{\textrm{CO}}$), predict a steep power-law relationship between $\alpha_{\textrm{CO}}$ and the gas metallicity of galaxies (e.g., \citealt{Israel1997}). Should their predictions hold true, many high-redshift galaxies may lack significant CO emission, despite the presence of molecular gas \citep{Wolfire2010,Munoz2014}. Other theoretical work (e.g., \citealt{Glover2011,Obreschkow2009b}) suggests that CO is not so strongly affected by the lower metallicity and dust masses of early galaxies, offering a more optimistic outlook for CO as a tool for exploring molecular gas at high redshift.

The makeup of the molecular gas content of star-forming galaxies in the early Universe is currently an active area of observational research. Recently, \cite{Decarli2014} and \citet{Walter2014} used the Plateau de Bure Interferometer (PdBI) to make a very deep (100 hour integration time) observation of a portion of the GOODS-N field \citep{Dickinson2003}, and probed the CO luminosity function at $z\sim3$ to a limit of $L^{\prime}_{\textrm{CO}}\gtrsim 10^{10}\ \textrm{K}\ \textrm{km}\ \textrm{s}^{-1}\ \textrm{pc}^{2}$ for the $J{=}3{\rightarrow}2$ rotational transition of CO. For $z\sim3$, this line luminosity limit corresponds to galaxy SFRs greater than $\sim 1.2\times10^{2}\ M_{\odot}\ \textrm{yr}^{-1}$ \citep{Tacconi2013}. Clearly, such studies are limited to massive galaxies that are rapidly forming stars and miss the lower luminosity and/or less massive systems that are expected to make up a large fraction of star-forming galaxies \citep{Obreschkow2009b,Lagos2011,Smit2012,Bouwens2012,Sargent2014}.

Exploration of the properties of more typical galaxies may be done through ``intensity mapping'', where the signals from hundreds or thousands of galaxies -- both bright and dim -- are detected in aggregate as larger-scale fluctuations in the mean line intensity. The method of CO intensity mapping has been investigated in numerous recent theoretical studies \citep{Righi2008,Visbal2010,Carilli2011,Visbal2011,Gong2011,Lidz2011,Pullen2013,Li2015}. These analyses have predicted the mean brightness temperature of CO at $z\sim3$ to be of order $\langle T_{\textrm{CO}} \rangle \sim 1\ \mu\textrm{K}$ -- within reach of existing instruments with very deep integrations, provided that observational systematics can be controlled.

The Sunyaev-Zel'dovich Array (SZA), an 8-element interferometer of 3.5 meter dishes with 1 cm receivers, is well matched to performing a CO intensity mapping experiment. The SZA is capable of observing the $J{=}1{\rightarrow}0$ rotational transition of CO, which we will herein refer to as $\textrm{CO}(1{-}0)$, at a redshift range of $z=2.3{-}3.3$, with greatest sensitivity to comoving size scales of $0.5{-}2\ h^{-1}\,\textrm{Mpc}$. The data set for our analysis was previously used by \citet{Sharp2010} (hereafter \citetalias{Sharp2010}) as a measurement of the cosmic microwave background (CMB) power spectrum on arcminute angular scales. 

In this paper, we discuss our search for measurable anisotropy in the three-dimensional distribution of molecular gas by characterizing the variance spectrum of the \citetalias{Sharp2010} data. This analysis resembles the two-dimensional power spectrum measurement of that work, extended to a third dimension using the frequency channels recorded in the data. Sensitive measurements may eventually image the intensity variations due to the large-scale structure of CO-emitting galaxies, but at the depth of these data we expect, at best, only a statistical detection of CO fluctuations via variance that exceeds that expected from the thermal noise of the data set.

This paper is structured as follows: Section~\ref{sec_observations} discusses the SZA instrument and the observational data used for this experiment. Section~\ref{sec_analysis} discusses the software pipeline built for the analysis of the data. Section~\ref{sec_results} presents the results of our analysis, and Section~\ref{sec_discussion} discusses these results in the context of theoretical expectations. Conclusions are given in Section~\ref{sec_conclusion}.

\section{Observations}\label{sec_observations}
In this section, we present a brief description of the instrument and data set used in our analysis. A more thorough description of each can be found in \cite{Muchovej2007} and \cite{Sharp2010}, respectively.
\subsection{Instrument Description}\label{ssec_arraydescrip}
\begin{figure*}
\begin{center}
\includegraphics[scale=0.5]{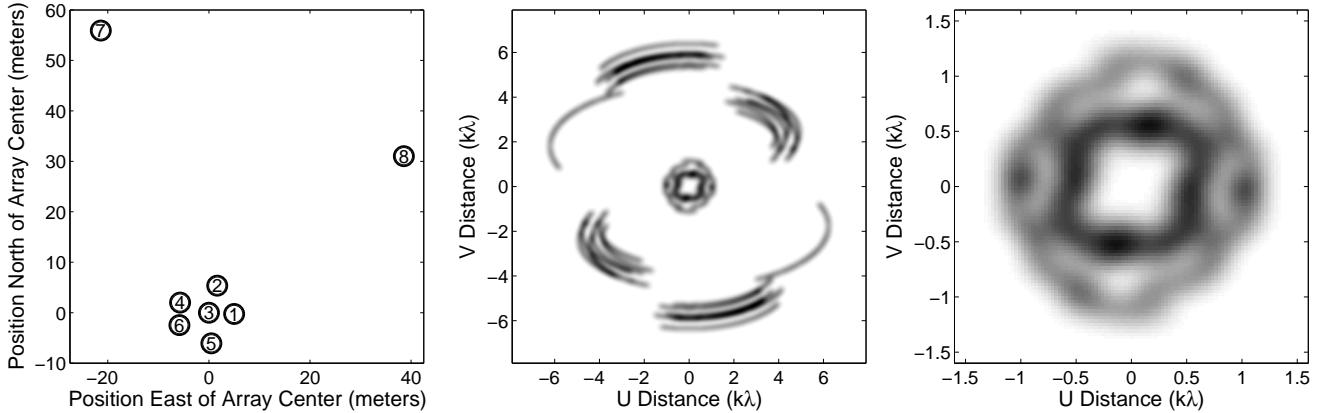}
\caption{\textit{Left}: SZA antenna positions. The antennas are drawn as 3.5 meter diameter circles. 
\textit{Center:} the $uv$ coverage of the full array for one spectral window (30.2 GHz) for one of the fields in the \citetalias{Sharp2010} survey, with darker shades corresponding to areas with greater sensitivity. The baselines to the outrigger antennas, 7 and 8, are well-separated from the baselines within the compact portion of the array. 
\textit{Right}: The inner region of the $uv$ plane.
\label{fig_uvcoverage}}
\end{center}
\end{figure*}
The Sunyaev-Zel'dovich Array (SZA) is an 8-element interferometer designed for measurements of the Sunyaev-Zel'dovich (SZ) effect signature of galaxy clusters on arcminute angular scales \citep{Muchovej2007}. At the time of data collection, the SZA was located at the Owens Valley Radio Observatory (OVRO), it was later incorporated into the nearby Combined Array for Research in Millimeter-wave Astronomy. 

Each SZA antenna has a single polarization (left-hand circular) 1-cm receiver, capable of observing between $27{-}35\ \textrm{GHz}$, which corresponds to $z\approx 2.3{-}3.3$ for CO$(1{-}0)$. At 30 GHz, an SZA antenna has a typical system temperature $T_{\textrm{sys}}\approx 40\ \textrm{K}$, and a typical aperture efficiency of $\eta_{\textrm{eff}}\approx 0.6$. Each antenna is $3.5\ \textrm{m}$ in diameter, which corresponds to a FWHM $\theta_{\textrm{B}} \approx 11'$ at $30$ GHz and covers an effective solid angle of $\Omega_{\textrm{B}}=0.03\ \textrm{deg}^{2}$. The SZA antennas were arranged in a compact group of 6 antennas (4.5-11.5~m spacings), with two outrigger antennas $\sim$50 meters away to separate the extended SZ signal from compact radio continuum sources.  Figure~\ref{fig_uvcoverage} shows both the layout of the SZA as well as the $uv$ coverage of a single field during a typical observation.

The SZA operates using a 2-bit digital XF correlator -- with an efficiency of $\eta_{\textrm{corr}} = 0.87$ -- providing a total of 28 cross-correlations and 8 autocorrelations. The correlator has a total bandwidth of 8 GHz split across 16 different windows of 500 MHz bandwidth. Within each window are 16 spectral channels, of width 31.25 MHz. One channel is discarded within each window due to edge effects, leaving 468.75 MHz of usable bandwidth per window.

\subsection{Data Description}\label{ssec_datadescrip}

The \citetalias{Sharp2010} data were obtained between 2005 and 2008. The data consist of 44 telescope pointings, arranged as 11 groups of 4 fields. Each group is composed of 4 pointings at constant declination separated by 4 minutes in RA, such that each field is observed over the same hour angle over a series of sequential 4 minute observations. The duration of the loop through the 4 fields, consisting of twenty 20-second integrations per field, plus several minutes on a gain calibrator is approximately 20 minutes. A bandpass calibrator was typically observed for 5 minutes at the beginning or end of the track. We refer to a contiguous track of data, typically 6 hours in length, as an ``observing block''. Each group of 4 fields was observed for approximately 45 days, providing an hour a day of integration time, with an average total of 20 hours of integration time per field after taking into account data flagging. A listing of the position of the lead field for each group is provided in Table~\ref{table_fields}.
\begin{table}[b]
\begin{center}
\caption{A listing of the lead field for each of the 11 groups observed in the \citetalias{Sharp2010} data set.
\label{table_fields}}
\begin{tabular}{| c | c | c | c |}
\hline
Field Name & RA & Dec & Gain Cal \\
\hline
cmbA1 & 02$^{\textrm{h}}$12$^{\textrm{m}}$00$^{\textrm{s}}$.0 & +33$\degr$00$'$00$''$ & J0237+288 \\
cmbAA1 & 21$^{\textrm{h}}$24$^{\textrm{m}}$38$^{\textrm{s}}$.7 & +25$\degr$29$'$37$''$ & J2139+143 \\
cmbBB1 & 21$^{\textrm{h}}$24$^{\textrm{m}}$38$^{\textrm{s}}$.1 & +25$\degr$59$'$24$''$ & J2025+337 \\
cmbCC1 & 02$^{\textrm{h}}$11$^{\textrm{m}}$31$^{\textrm{s}}$.3 & +33$\degr$27$'$43$''$ & J0237+288 \\
cmbDD1 & 13$^{\textrm{h}}$18$^{\textrm{m}}$40$^{\textrm{s}}$.1 & +35$\degr$01$'$42$''$ & J1131+305 \\
cmbEE1 & 14$^{\textrm{h}}$18$^{\textrm{m}}$39$^{\textrm{s}}$.2 & +35$\degr$31$'$52$''$ & J1331+305 \\
cmbI1 & 02$^{\textrm{h}}$12$^{\textrm{m}}$00$^{\textrm{s}}$.0 & +32$\degr$37$'$08$''$ & J0237+288 \\
cmbR1 & 02$^{\textrm{h}}$12$^{\textrm{m}}$15$^{\textrm{s}}$.6 & +32$\degr$11$'$24$''$ & J0237+288 \\
cmbY1 & 02$^{\textrm{h}}$12$^{\textrm{m}}$00$^{\textrm{s}}$.0 & +31$\degr$51$'$24$''$ & J0237+288 \\
cmbXX1 & 21$^{\textrm{h}}$24$^{\textrm{m}}$38$^{\textrm{s}}$.7 & +24$\degr$59$'$37$''$ & J2139+143 \\
cmb07 & 02$^{\textrm{h}}$07$^{\textrm{m}}$37$^{\textrm{s}}$.0 & +34$\degr$00$'$00$''$ & J0237+288 \\
\hline
\end{tabular}
\end{center}
\end{table}
\section{Analysis}\label{sec_analysis}
\subsection{Pipeline Overview}\label{ssec_pipeline}
This section provides a broad overview of the data processing and calibration software used in our analysis. The power spectrum analysis and null tests are described in Sections~\ref{ssec_powspecgen} and~\ref{ssec_jackknife}, respectively. All of the routines used here were developed within MATLAB\footnote{Mathworks, Version 2013b, \url{http://www.mathworks.com/products/matlab/}}. 

The calibration of the raw SZA data follows similar procedures to those described in \citet{Muchovej2007} and \citetalias{Sharp2010}. The raw data are recorded as complex correlation amplitudes with associated time-tagged status information. They are converted to a physical power scale using system temperature measurements that are made during every source-calibrator cycle. Absolute telescope and system efficiencies, derived from Mars via the \citet{Rudy1987} brightness temperature model, are applied. As these factors are identical to those used in \citetalias{Sharp2010}, we expect they are accurate to 10\%. The data are flagged to remove bad data, bandpass calibration is determined from a strong point source in each track, and relative gain calibration is determined from the gain calibrator observed on a 20-minute cycle. Some features of these steps are outlined below. 

Flagging of data is done using three principal methods. First, data affected by known hardware issues (e.g., an antenna fails to point correctly) are marked as bad. Second, data are passed through various statistical tests and checked against theoretical estimates to see if they behave in a Gaussian fashion (data that exceed estimated noise thresholds are flagged as bad). Finally, data adjacent (i.e., belonging to preceding or subsequent frequency channels or integrations) to bad data are removed as well. Flagging of data is typically done with an iterative approach, identifying outliers in small groups of data (and removing them when appropriate) before reevaluating the data over larger groupings. This approach helps limit the impact that a few bad data points may have on an otherwise good data set, at the expense of additional processing time. We remove 28.7\% of the data for known hardware problems (including shadowing of antennas), 4.8\% of data for exceeding noise thresholds, and 6.5\% of data for bad neighboring data. We also discard all data from our shortest baseline (amounting to 3.6\% of our data additionally removed) due to known systematic noise issues. In total, 43.6\% of all data are flagged.

The SZA system produces very stable calibration across frequency and time. Gain amplitude and phase typically vary by 2\% and $<10^\circ$ across a track, respectively. When large gain shifts are observed ($>10\%$ in amplitude or $>30\degr$ in phase) between gain calibrator observations, data are marked as bad and excluded from later analysis. Phase solutions are linearly interpolated, while the very stable gain amplitudes are averaged over the track. Bandpass solutions are also typically stable to 1\% between observing blocks.

\begin{figure}[!t]
\begin{center}
\includegraphics[scale=0.5]{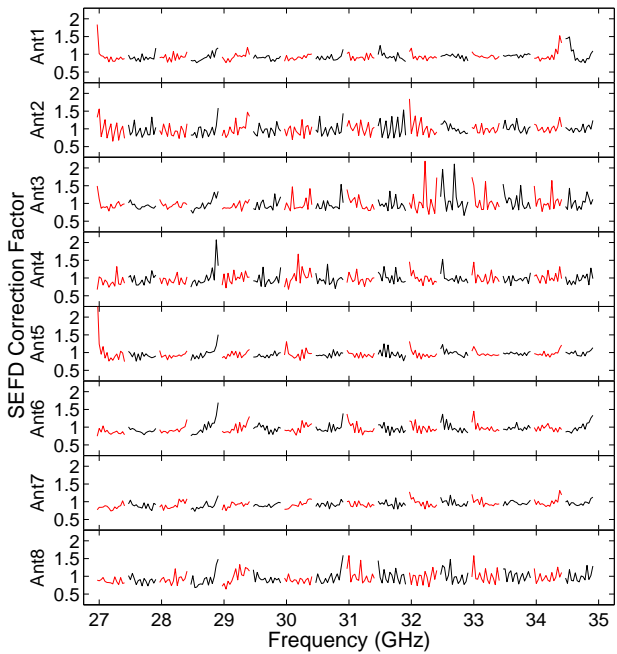}
\caption{An example of the SEFD correction solution generated for BB20070906 -- a single observing block within our data set. SEFD corrections show good day-to-day agreement, similar to the bandpass calibration solutions.
\label{fig_noiseexample}}
\end{center}
\end{figure}

One additional calibration step is performed to account for discrepancies between the expected and actual noise within our measurements. System temperatures are measured for each window, but RF/IF features and quantization effects can introduce variability in the system temperature on a channel-by-channel basis. To account for this effect, the variance within each channel of each baseline is calculated for all data within a single observing block (excluding calibrator data, and after subtraction of known sources in each field), and the ``system equivalent flux density (SEFD) correction'' is determined as the difference between the theoretical and measured noise. These differences are believed to be due to antenna-based effects (primarily due to  
standing waves in the receiver bandpass that introduce a spectrally-varying signal level at the digitizers, resulting in varying quantization noise) hence an antenna-based correction factor is determined (using a $\chi^{2}$ fit) for each frequency channel. The correction factors are determined once per track and are seen to be consistent to 1\% between observing blocks. An example SEFD correction spectrum is shown in Figure~\ref{fig_noiseexample}.
\begin{figure*}[!t]
\begin{center}
\includegraphics[scale=0.575]{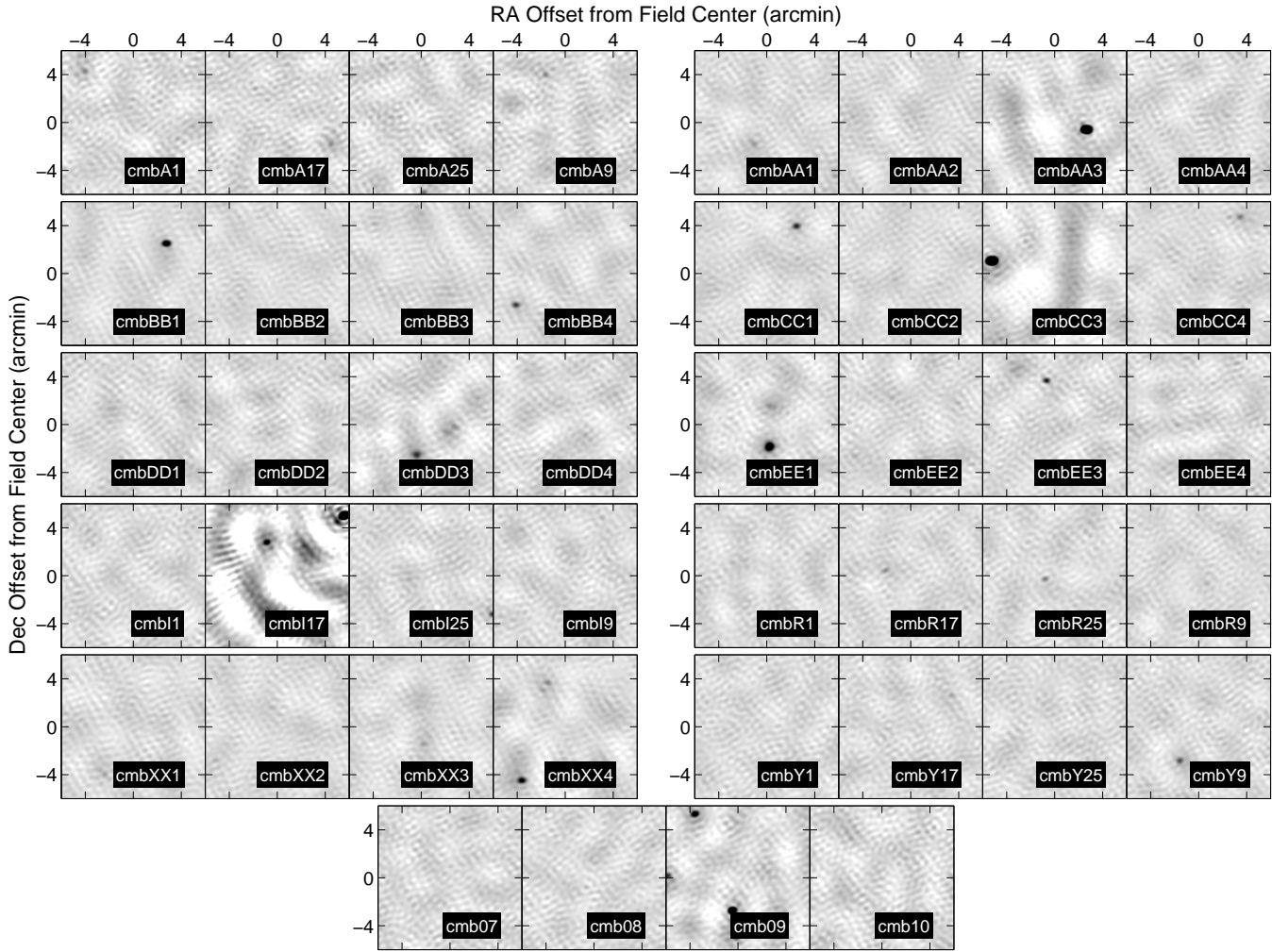}
\caption{Thumbnail images of the deconvolved maps for all 44 fields included in the \citetalias{Sharp2010} survey. On average, we find there to be a rough average of one bright point source ($S\gtrsim1\ \textrm{mJy}$) per pointing. Most fields after deconvolution show noise-like residuals, though cmbI17 shows some artifacts due to the presence of $100\ \textrm{mJy}$ point source at the edge of the primary beam -- arguably the worst-case scenario for contamination of our measurements (see Section~\ref{ssec_jackknife}).
\label{fig_thumbimages}}
\end{center}
\end{figure*}

Figure~\ref{fig_thumbimages} presents images of all 44 fields, deconvolved via the CLEAN algorithm \citep{Hogbom1974}. No primary beam correction is applied to these images. The resolution of the images is $\sim0.5'$ and the median theoretical noise is $0.15\ \textrm{mJy}$. We find that the measured RMS residual noise in our images following deconvolution is consistent with theoretical estimates based on thermal noise, with the exception of cmbI17 (due to the presence of a bright point source at the field edge). We detect 24 sources above 10$\sigma$ across the 44 fields, with a median flux of 2.5 mJy and a maximum (primary beam corrected) flux of 100 mJy. These sources are subsequently removed from our data prior to our power spectrum analysis. Sources below the $10\sigma$ threshold are ignored because of their limited impact on our power spectrum measurement (see Section~\ref{ssec_simulations}). These results are consistent with those reported by \cite{Sharp2010}.

\subsection{Power Spectrum Analysis}\label{ssec_powspecgen}
The goal of our power spectrum analysis is to transform the frequency-resolved interferometric visibilities into a measurement of the CO power spectrum. We provide an overview of the method here, reserving the full details for Appendix~\ref{app_maptechnique}. 

The power spectrum measurement is performed in a three-dimensional spatial frequency domain that is the Fourier transform of the real-space universe. The interferometer naturally provides measurements of spatial frequencies in the two dimensions of the sky, while the third is constructed by Fourier transforming the frequency dimension of the data, which maps to the line of sight distance via the redshift of the CO line. The data are Fourier transformed one 500 MHz window at a time, providing redshift-segmented measurements of the Fourier-transformed intensity field, $\tilde{I}(u,v,\eta,z)$, with $z$ representing the median redshift of the window. We refer to these as ``delay-visibilities.'' The delay-visibilities are then converted to temperature units, weighted by their variance estimates and gridded in the $(u,v,\eta,z)$ space. Here $u$ and $v$ are the standard spatial frequency variables for interferometry, and $\eta$ represents the Fourier transform of the frequency axis. The $(u,v,\eta)$ domain is closely related to the $\mathbf{k}$ space, differing only by the conversion factors $X$ and $Y$,  which convert between comoving physical size and angular distance or frequency with units $\textrm{Mpc}\ \textrm{rad}^{-1}$ and $\textrm{Mpc}\ \textrm{Hz}^{-1}$, respectively \citep[e.g.,][]{Parsons2012}. With the application of the $X$ and $Y$ conversion factors, we find the Fourier dual of the specific intensity in the three-dimensional $k$-vector space, $\tilde{I}(\mathbf{k})$. Within an individual window our coverage of $\mathbf{k}$ space is similar in the $\eta$ direction and the $u$ and $v$ directions, considering only the compact portion of the array.

Formally, the intensity power spectrum is defined as
\begin{eqnarray}\label{eqn_powspec}
\Delta^{2}(k) &\equiv& \frac{k^{3}}{2\pi^{2}} P(k), \nonumber \\
&=& \frac{k^{3}}{2\pi^{2}} \frac{X^{4}Y^{2}}{\textrm{V}_{z}}\left(\frac{c^{2}}{2k_{B}\nu^{2}}\right)^{2} \left \langle \left | \tilde{I}^{2}\right | \right \rangle_{\mathbf{k}\cdot\mathbf{k}=k^{2}}.
\end{eqnarray}
In Equation~\ref{eqn_powspec}, $\Delta^{2}(k)$ (in units of $\mu\textrm{K}^{2}$, often called the dimensionless power spectrum; e.g., \citealt{Furlanetto2006}) is the variance in brightness temperature at comoving spatial frequency $k$ (in units of $h$~Mpc$^{-1}$, where $h=H_{0}/100\ \textrm{km}/\textrm{s}/\textrm{Mpc}$ and $H_{0}$ is the current Hubble parameter) per ln($k$). The power spectrum, $P(k)$, is the Fourier transform of the autocorrelation function of the CO intensity field, and has units $\textrm{Mpc}^{3}\ \mu\textrm{K}^{2}$. The Boltzmann factor is represented by $k_{B}$, and the speed of light by $c$. The volume probed by the measurement at redshift $z$ is $\textrm{V}_{z}=X^{2}YB_{z}\Omega_{\textrm{B}}/2$, for a solid angle of the telescope primary beam, $\Omega_{\mathrm{B}}$, and a bandwidth, $B_{z}$, considered around this redshift. For a single pointing, the volume surveyed by a single correlator window (in this case, the central one) is a cylinder of diameter 12 $\textrm{Mpc}/h$ and length 45 $\textrm{Mpc}/h$.

As a dimensionless quantity, $\Delta^{2}(k)$ is sometimes favored over $P(k)$ \citep{Dodelson2003}. However, $P(k)$ has the advantage that it maintains the same value for different values of $k$ when in the cosmological shot noise limit; for both the CO measurement and some foreground contaminants, our experiment resides firmly in the shot noise regime. As such, values in both conventions are presented throughout this paper.

Values for the power spectrum are calculated using the following equation:
\begin{eqnarray}
\mathcal{P}(\mathbf{k},z) &=& \frac{\sum\limits_{\mathbf{k}^\prime}\sigma_{k}^{-2}\sigma_{k^{\prime}}^{-2} C(\mathbf{k}-\mathbf{k}^{\prime})\left( | \tilde{I}^{*}(\mathbf{k},z)\tilde{I}(\mathbf{k}^{\prime},z) |\right)}{\sum \limits_{\mathbf{k}^\prime} \sigma_{k}^{-2} \sigma_{k^{\prime}}^{-2} C^{2}(\mathbf{k}-\mathbf{k}^{\prime})} - \mathcal{A}_{\mathbf{k}} \nonumber \\
P(k,z) &=& \left \langle \mathcal{P}(\mathbf{k},z) \right \rangle_{\mathbf{k}\cdot\mathbf{k}=k^{2}} \label{eqn_powspecavg}.
\end{eqnarray}
In Equation~\ref{eqn_powspecavg}, gridded data within a single redshift window are cross-multiplied against one another, weighted by their estimated thermal noise variance, $\sigma_{k}^{2}$, and their normalized covariance, $C$, for the cross-multiplied data. The covariance is the analytically calculated correlation between adjacent points in the $(u,v,\eta)$ space. For the $uv$ plane, this amounts to a consideration of the overlap of the visibilities being multiplied, each of which samples an area in the $uv$ plane described by the autocorrelation of the telescope illumination pattern. The correlation is similarly calculated between $\eta$ channels. To eliminate noise bias from our measurement, the autocorrelations of individual delay-visibilities are averaged over each grid cell ($\mathcal{A}_{\mathbf{k}}$) and removed from the power spectrum measurement. As a result, the autocorrelation measured in our power spectrum analysis may produce negative values when the result is noise-dominated.  The power spectrum measurement is collapsed down to a single dimension, averaging over spheres in the $k$ space. A power spectrum is created for each redshift window, within each source field.

To test the fidelity of the power spectrum pipeline,  we generate a series of simulated observations of mock fields. Visibilities are generated based on a randomized set of spectral line sources. Gaussian noise is added to each visibility using noise estimates based on system temperature, instrument parameters and SEFD corrections. Effects such as primary beam attenuation and mode mixing (i.e., bandwidth smearing) are taken into account in these simulations, to better understand the limits of our analysis. To account for shot noise, we simulate data for $10^{4}$ fields, which are then averaged together to provide a final power spectrum result. We find that the input power is recovered across our full $k$ range to $1\%$, limited by small spectral and gridding corrections.

\subsubsection{Ground Subtraction}
The lead-trail design of the \citetalias{Sharp2010} observations allows the removal of any contributions to the power spectrum that are correlated with antenna position, such as ground contamination or antenna cross-talk. It was found in \citetalias{Sharp2010} that without ground subtraction there was weak evidence for a ground-correlated contamination of the CMB power spectrum measurement. Similar tests, described in the next section, suggest that removal of such contamination is important to cleaning our data. 

We create a model for the contamination using a variance-weighted average of the four fields in each group. The average is generated visibility by visibility, so that the first 20-second integrations on each field in the group are averaged together, preserving the individual baselines and frequency channels, as are the second integrations, and so on. This model is subtracted from the individual visibilities in each of the four fields, reducing the number of independent measurements in our experiment by 25\%, which degrades the sensitivity by approximately 12\%. 

\subsubsection{Point Source Contamination}\label{ssec_simulations}
\begin{figure}[b]
\begin{center}
\includegraphics[scale=0.5]{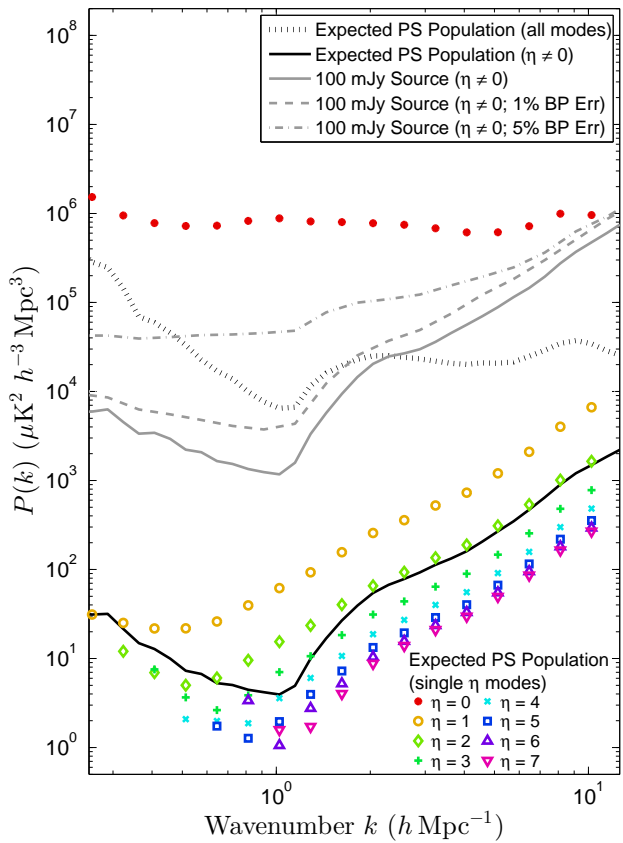}
\caption{Results of simulations to probe the impact of point sources on our analysis. Separate simulations were used to probe the impact of a bright 100~mJy source (gray solid) and of the expected population of point sources. The measured power, averaged over three-dimensional shells in $k$, is shown before (black dotted) and after (black solid) removal of the $\eta=0$ modes. For the 100~mJy source simulations, we additionally show the impact that bandpass errors have on the measured residual power, for RMS errors of 1\% (gray dashed) and 5\% (gray dot-dashed). For the expected point source population simulations, we additionally show the power measured when using data from a single $\eta$ channel.
\label{fig_pspowspec}}
\end{center}
\end{figure}
Continuum point sources have the potential to significantly contaminate our measurements. Point sources provide of order $10^{3}$ more power (in units of $\mu\textrm{K}^{2}$) than CMB anisotropies or Galactic foregrounds over the range of multipoles ($\ell\sim3000$) that the SZA is sensitive to (\citepalias{Sharp2010}; \citealt{Bennett2013,PlanckXI2015}). As such, we expect point sources to be the dominant astrophysical contaminant in our measurements. Fortunately, most sources at 30 GHz will be close to spectrally flat over the frequency range of one window \citep{Muchovej2010}, and therefore will contribute power primarily to Fourier modes with $\eta=0$. To reject this contamination, we drop the $\eta=0$ channel. Dropping this channel does not significantly affect the $k$ range the experiment, as the length of the smallest three dimensional $k$ vector is determined primarily by the shortest baselines, rather than the smallest $\eta$ mode.

We test the effectiveness of our method of rejecting continuum source power with three sets of simulations. First, we simulated observations of mock fields with a 100 mJy point source of varying spectral index, inserted at a random position within twice the FWHM primary beam. Such a source is much brighter than average, and is somewhat brighter than the brightest single source found in our data. Second, we simulated observations of mock fields following the 30~GHz point source counts and spectral index distributions measured by the SZA in \cite{Muchovej2010} (which we refer to as the ``expected point source population''). We simulated sources of flux density up to 15~mJy, above which observations show a steep drop in point source counts. Third, we introduced calibration errors (i.e., baseline length, gains and bandpass) into both our expected point source population and 100 mJy source simulations. 

The results of these simulations are shown in Figure~\ref{fig_pspowspec}. For simulations probing the impact of the expected point source population, we additionally present a breakdown of the power measured in Fourier modes for individual $\eta$ channels. Power from continuum sources is predominantly found in the $\eta=0$ channel; as the one-dimensional power spectrum is an average over all remaining $\eta$, the effect of contamination is an average of the $\eta$ channels shown.

In the absence of calibration errors, we find that rejection of the $\eta=0$ mode reduces the contribution of continuum sources by factors of $10^{1}$ to $10^{4}$, depending on $k$. We find that the residuals rise roughly as $k^{2}$ for $k \gtrsim 2 \ h\,\textrm{Mpc}^{-1}$. This is due to the appreciable change of the baseline length over the bandwidth of a window for the long baselines of the array -- an effect sometimes referred to as `mode-mixing' \citep{Datta2010,Morales2012}. Our experiment does not achieve interesting sensitivity to these modes, so we do not expect that this effect requires more thorough mitigation. We find that rejection of the $\eta=0$ channel reduces the power spectrum contributions of a 100 mJy point source to levels comparable to the sensitivity of our experiment. In the real data, the suppression of power from bright point sources should be greater, as we remove such sources from the visibilities before calculating power spectra. We also find that rejection of the $\eta=0$ channel reduces the power spectrum contributions of the expected source population to levels well below our sensitivity threshold. Further testing reveals that the point source residual power decreases by an additional factor of $\sim10$ after removing sources of flux density greater than 1.5 mJy (roughly corresponding to the flux limit for source subtraction as discussed in Section~\ref{ssec_pipeline}).

In simulations that include calibration errors, we find that bandpass errors can significantly increase the residual power from continuum point sources. We find that for RMS fractional bandpass errors, $\sigma_{\textrm{BP}}$, between 1\% and 10\%, the residual power scales as $\sigma_{\textrm{BP}}^{2}$. We expect that our bandpass errors are only 1\% (discussed in Section~\ref{ssec_pipeline}); our simulations indicate that this will increase point source residuals by 30-300\%, depending on $k$. This effect is negligible for the expected point source population, though it may allow residuals from bright point sources to exceed our detection threshold. We note these simulations represent a worst-case scenario, as only a single set of bandpass errors are incorporated into each mock field. In real data, a new bandpass solution is derived for each observing block; hence, data for a typical field implements 45 different bandpass solutions. Assuming that the errors in these bandpass solutions are thermal-noise dominated, we expect the increase in residual power to be only a few percent of what we have found in these simulations.

We further find that gain and baseline errors do not significantly increase the residual point source power, though they do impact the fidelity with which we are able to remove bright sources from our data set. We also note that calibration errors of all varieties will reduce the sensitivity of our measurement (as these errors will ``wash out'' the signal of interest), by roughly a factor equal to the fractional error in the calibration. Given that our calibration errors are of order one part in a hundred, we expect this loss of sensitivity is likely to be minimal.
\subsection{Jackknife Tests}\label{ssec_jackknife}
To determine the impact of systematic errors on our measurements, we conduct a set of null tests in which we remove the astronomical signal from our data via linear combination or randomization and search for residual power. These tests, commonly referred to as jackknife tests, are split into four different categories: intraday, interday, ``cross-window'' and noise tests. Results from our jackknife analysis can be found in Section~\ref{ssec_powspec}.

Intraday tests take temporally adjacent visibilities (i.e., subsequent integrations of the same channel within the same baseline) and modulates them in such a way that the sky signals are canceled out. Two tests are performed in this manner: ``couplet'' tests (visibilities adjacent in time in a given baseline are simply subtracted from one another) and ``triplet'' tests (three adjacent visibilities are multiplied by a phase offset of $0\degr$, $120\degr$, and $240\degr$ respectively before being summed together). Visibilities are gain-corrected and flagged prior to this operation. In the event that one visibility of the couplet/triplet group is flagged as bad, the entire group is thrown out and not considered in subsequent jackknife analysis. Both couplet and triplet tests are sensitive to high cadence systematics within our data set, with the couplet and triplet tests evaluating the data set over respectively shorter and longer periods.

Interday tests utilize gridded data rather than visibilities. Data from different observing blocks (i.e., days) are summed into two separate stacks and subtracted from one another. Two different interday tests are performed. The ``even-odd'' test sorts observations from alternating days into two stacks, and then subtracts them. The ``first-last'' test sums together data from all days belonging to the first half of the set of days and subtracts them from the sum of the second half.

The cross-window test operates by correlating different redshift windows with one another. The CO signal is not expected to correlate between different redshift windows, hence no subtraction-like step is necessary. However, continuum contamination that is not rejected by the removal of the $\eta=0$ channel will correlate between windows, so this jackknife is an empirical check on our continuum rejection.

To further test the analysis software and pipeline, we perform a pair of tests to measure and confirm our noise estimates. In both tests, values are multiplied by a random complex number with absolute value of 1 and random phase. These ``randomized phase'' tests gives us our best verification of noise estimates, as any potentially coherent emission detected by the interferometer should be scattered and thus not affect the results of this test. The ``random raw'' test randomizes the phases of ungridded data, while the ``random grid'' randomizes the phases of gridded data. While the former is much more thorough, the latter requires much less processing time to complete (while still being a useful tool for the verification of noise estimates).

\begin{table}[t]
\begin{center}
\caption{Table of jackknife test results
\label{table_results}}
\begin{tabular}{|c|c|c|}
\hline
Result & Measured $P$ and $1\sigma$ errors &\\ 
Type &  $ (10^{4}\ \mu\textrm{K}^2\ h^{-3}\,\textrm{Mpc}^{3})$ & PTE\\
\hline
Couplet & $2.00 \pm 1.74$ & 0.25 \\
Triplet & $1.21 \pm 2.46$ & 0.62 \\
Even-Odd & $1.16 \pm 1.34$ & 0.39 \\
First-Last & $0.22 \pm 1.47$ & 0.88 \\
Cross-Win & $-0.16 \pm 1.60$ & 0.92 \\
\hline
\end{tabular}
\end{center}
\end{table}

\begin{figure}[t]
\begin{center}
\includegraphics[scale=0.5]{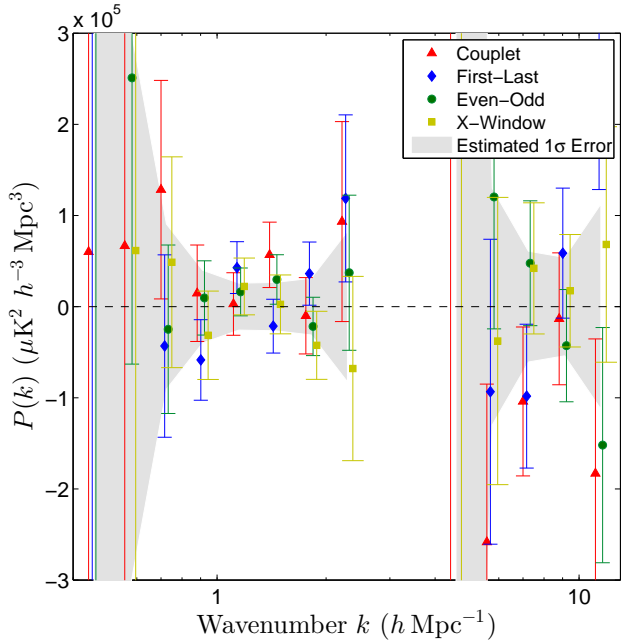}
\caption{Jackknife results for our analysis of the \citetalias{Sharp2010} data set, alongside the $1\sigma$ noise threshold for our measurement (solid gray). For each bin, we find the results to be noise-like in their distribution, consistent with our assumption that the data are predominately free of systematics that may affect our result. Our largest outlier has a significance of $2.0\sigma$, consistent with what one would expect for a set of $\sim60$ values that have been normally distributed.
\label{fig_jackknife}}
\end{center}
\end{figure}

We show the results of our jackknife analysis in Figure~\ref{fig_jackknife} and in Table~\ref{table_results}. Listed in the table are the measured values (and associated errors) for each of the tests. Additionally listed for each test is the probability to exceed (PTE) -- the likelihood for a noise-like event to produce a measurement of equal or greater statistical significance. Our jackknife results appear to be consistent with noise, suggesting that our analysis is not dominated by systematics. We note that our cross-window test correlates windows that are two steps apart (e.g., window 1 with window 3, window 2 with window 4). When correlating adjacent windows with one another (e.g., window 1 with window 2, window 3 with window 4), the cross-window test finds power at a level of $2.3\sigma$ significance. We attribute this power to signal leakage between windows (brought on by minor imperfections in analog bandpass filtering), and do not expect it to impact our analyses.

Additionally, we find good agreement between estimates for the RMS power induced by thermal noise and the random grid test. The RMS power is estimated by utilizing the measured system temperatures, aperture efficiency, and SEFD corrections to estimate the amount of noise power in a given delay-visibility (and propagating that noise estimate forward using Equation~\ref{eqn_powspec} to estimate the noise power in the power spectrum). The calculated uncertainty and the uncertainty derived from the random grid test agree to 10\%, as we would expect for the 100 trials used in the test. We do find some minor differences between the two estimates in bins with lowest sensitivity; these bins have a small number of independent modes used to measure the power in the bin (hence their errors will not be normally distributed), therefore some discrepancy is to be expected.

\section{Results}\label{sec_results}
\subsection{CO Power Spectra}\label{ssec_powspec}
\begin{figure*}[!t]
\begin{center}
\plottwo{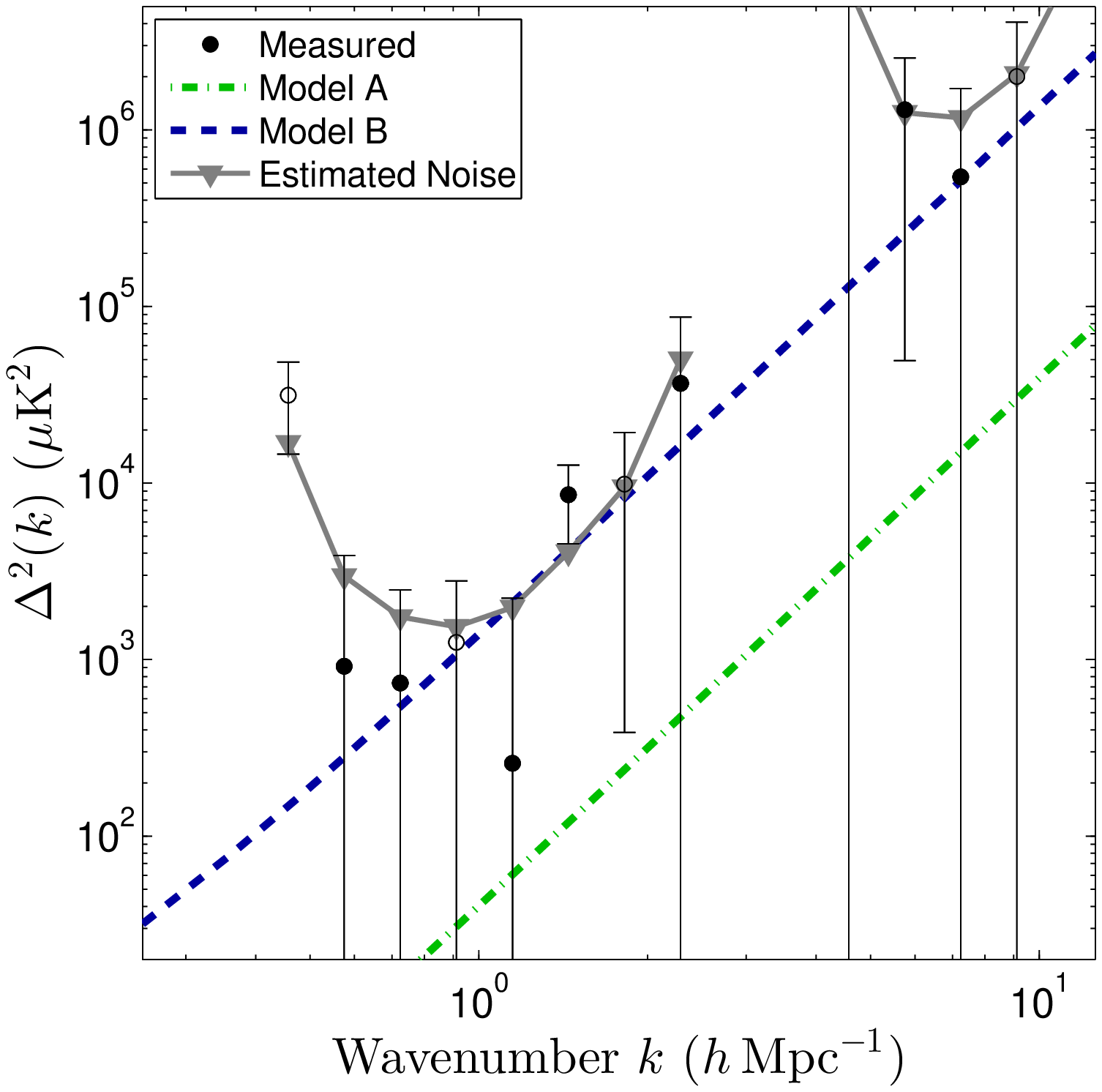}{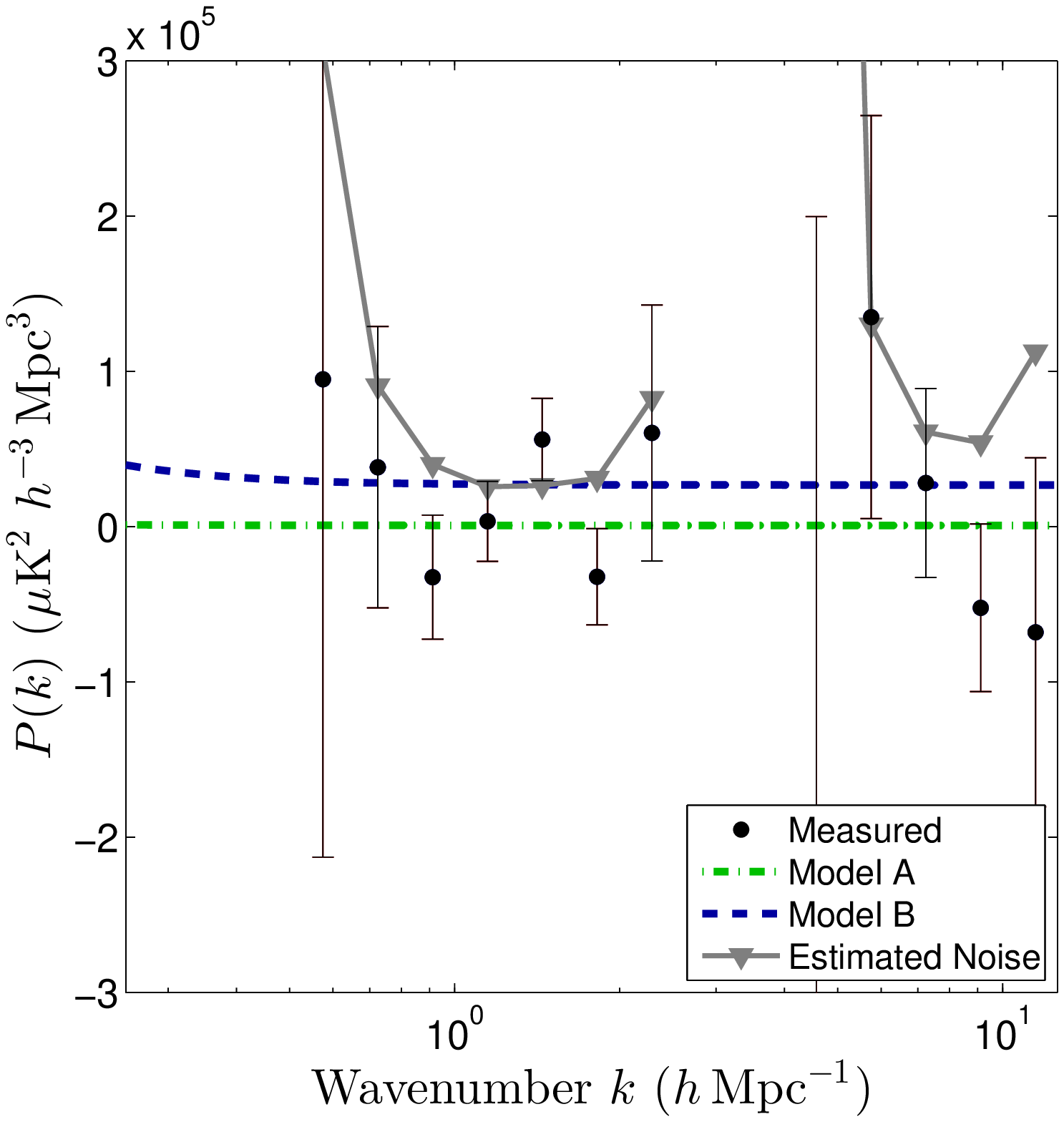}
\caption{\textit{Left}: The power spectrum result from our analysis of the \citetalias{Sharp2010} data, in the form $\Delta^{2}(k)$. Filled circles correspond to positive values for $\Delta^{2}(k)$, while open circles correspond to negative values, and the error bars corresponding to the $1\sigma$ errors on our measured values. There exists a gap in sensitivity around $k\sim4\ h\,\textrm{Mpc}^{-1}$, owing to the separation between baselines to the outrigger antennas and baselines within the compact portion of the array (see Figure~\ref{fig_uvcoverage}). For reference, model A (dot-dashed green) and model B (dashed blue) from \cite{Pullen2013} are shown (discussed further in Section~\ref{ssec_comodels}), along with the estimated RMS noise power (gray triangle), absent any astrophysical signal. \textit{Right}: The power spectrum result, in the form of $P(k)$.}
\label{fig_finpowspec}
\end{center}
\end{figure*}
Presented in Figure~\ref{fig_finpowspec} are the final results of our analysis of the \citetalias{Sharp2010} data set. Our measurement has peak sensitivity at $k \! = \! 1.3\ h\,\textrm{Mpc}^{-1}$, with best sensitivity between $k=0.5{-}2\ h \,\textrm{Mpc}^{-1}$. Integrating over all redshift windows and wavenumbers, we see no evidence for excess power from CO at $z\sim3$ above a $2\sigma$ noise threshold of $P_{N}=2.6\times10^{4}\ \mu\textrm{K}^{2} (h^{-1}\,\textrm{Mpc})^{3}$. Placing this measurement into $\Delta^{2}_{N}$ units, where Poisson power grows like $k^3$, requires choosing a $k$ value. At $k \! = \! 1\ h\,\textrm{Mpc}^{-1}$, $\Delta^{2}_{N}=1.3\times10^{3}\ \mu\textrm{K}^{2}$. 

\begin{figure}[t]
\begin{center}
\includegraphics[scale=0.5]{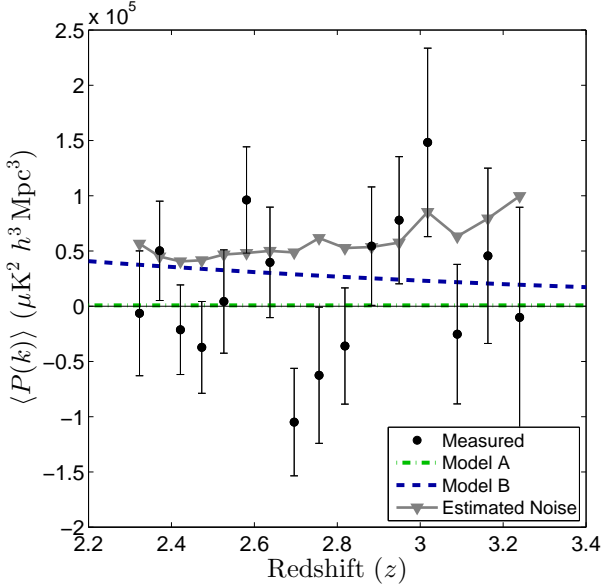}
\caption{Measured variation of $P(k)$, averaged over all $k$, as a function of redshift, with the $1\sigma$ error bars shown for each point. We find that the results are consistent with noise, with a maximum significance $2.0\sigma$ at $z=2.58$. The $2\sigma$ confidence upper limit resides just above the Model B prediction, with greatest sensitivity between $z=2.3{-}2.8$.
\label{fig_powspecz}}
\end{center}
\end{figure}

Theoretical models (e.g., \citealt{Pullen2013}) suggest that there may be significant evolution between the redshift range sampled by these data ($z=2.3{-}3.3$), thus we show in Figure~\ref{fig_powspecz} the results of our analysis for each redshift bin. With a mean $2\sigma$ noise limit of $P_{N}=10^{5}\ \mu\textrm{K}^{2}(h^{-1}\,\textrm{Mpc})^{3}$, $\Delta^{2}(k\!=\!1\ h\,\textrm{Mpc}^{-1})=5\times10^{3}\ \mu\textrm{K}^{2}$ within each redshift bin (of characteristic width $\Delta z = 0.06$), we see no evidence for excess power from CO within any of our individual redshift intervals.

\section{Discussion}\label{sec_discussion}
\subsection{Constraints on the CO Power Spectrum}\label{ssec_constraints}
\begin{figure}[t]
\begin{center}
\includegraphics[scale=0.5]{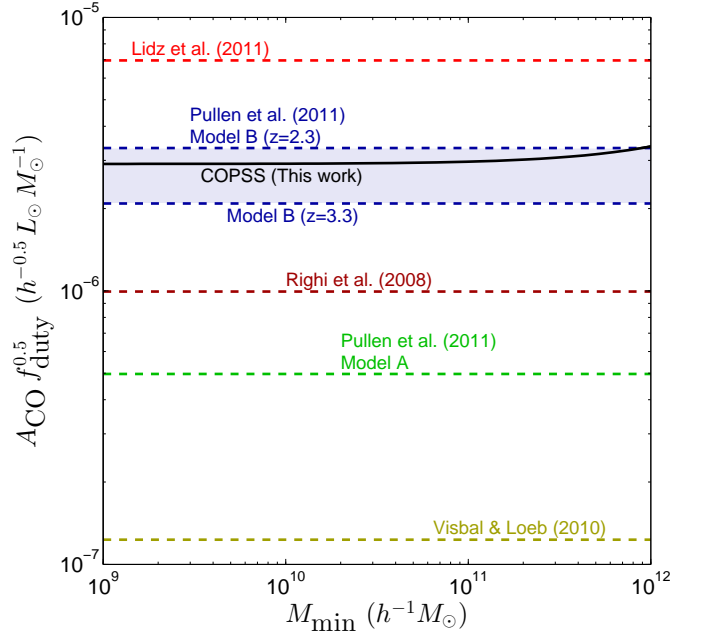}
\caption{Constraints on $A_{\textrm{CO}}$ as a function of $M_{\textrm{min}}$. The $2\sigma$ upper limit from our analysis (black solid) is shown versus several theoretical expectations for $A_{\textrm{CO}}$ \citep{Breysse2014}, multiplied by the square root of their adopted values for $f_{\textrm{duty}}$ ($f_{\textrm{duty}}=0.1$ for \citet{Visbal2011}, $f_{\textrm{duty}}=t_{s}/t_{H}$ for all others). \citet{Righi2008} do not explicitly supply a value for $A_{\textrm{CO}}$ or $f_{\textrm{duty}}$; we have therefore adopted the value of $A_{\textrm{CO}}$ calculated for this model by \citet{Breysse2014}, and have adopted the value for $f_{\textrm{duty}}$ from \citetalias{Pullen2013}.  The shaded region for model B reflects the model's variation with redshift over the range of our measurement.
\label{fig_mminvsa}}
\end{center}
\end{figure}
The power spectrum for CO as a function of wavenumber and redshift is given by
\begin{equation}\label{eqn_genpowspec}
P(k,z) = \langle T_{\textrm{CO}} \rangle^{2}b^{2}(z)P_{\textrm{lin}}(k,z) + P_{\textrm{shot}}(z),
\end{equation}
where $b(z)$ is the halo bias, $P_{\textrm{lin}}$ is the linear matter power spectrum and $P_{\textrm{shot}}$ is the shot-noise contribution to the power spectrum. One can further define $P_{\textrm{shot}}$ as
\begin{equation}\label{eqn_shotnoise}
P_{\textrm{shot}}(z) = \int \left( \frac{c^{3}(1+z)^2}{8\pi\nu_{o}^{3}k_{\textrm{B}}H(z)} L \right)^{2} \frac{dn(z)}{dL}\,dL,
\end{equation}
where $\nu_{o}$ is the rest frequency of the line (i.e., 115.271 GHz), $H(z)$ is the Hubble parameter, and $dn(z)/dL$ is the number of emitters per unit volume with luminosity $L$ at a given redshift $z$. If there exists a linear relationship between CO luminosity and halo mass (as many models assume; see Table 3 of \citealt{Li2015}), then Equation~\ref{eqn_shotnoise} can be written as 
\begin{multline}\label{eqn_shotnoisemass}
P_{\textrm{shot}}(z) =  \left ( A_{\textrm{CO}} \frac{c^{3}(1+z)^2}{8\pi\nu_{o}^{3}k_{\textrm{B}}H(z)} \frac{L_{\odot}}{M_{\odot}} \right)^{2} \\
f_{\textrm{duty}} \int_{M_{\textrm{min}}}^{\infty} M^{2} \frac{dn(z)}{dM} \,dM,
\end{multline}
where $A_{\textrm{CO}}$ is the ratio of CO(1-0) luminosity to host halo mass for CO-luminous halos, expressed in units of $L_{\odot}M_{\odot}^{-1}$  \citep{Lidz2011,Breysse2014}, $f_{\textrm{duty}}$ is the duty cycle of CO emitters, and $dn(z)/dM$ is the number of halos per unit mass as a function of redshift. Halos with masses below the low-mass limit, $M_{\textrm{min}}$, are assumed to be deficient in CO (either due to the suppression of star formation or because of a breakdown in the simple linear $M{-}L_{\textrm{CO}}$ relationship assumed), such that they do not appreciably contribute to $P_{\textrm{shot}}$ (e.g. \citealt{Pullen2013,Li2015}). Similarly, the mean brightness temperature can be expressed as
\begin{multline}\label{eqn_brighttemptoA}
\langle T_{\textrm{CO}} \rangle = A_{\textrm{CO}} \frac{c^{3}(1+z)^2}{8\pi\nu_{o}^{3}k_{\textrm{B}}H(z)} \frac{L_{\odot}}{M_{\odot}} \\
f_{\textrm{duty}}\int^{\infty}_{M_\textrm{min}}M\frac{dn(z)}{dM}\,dM.
\end{multline}
Using Equation~\ref{eqn_shotnoisemass} and provided with an appropriate halo mass function \citep{Tinker2008}, one can use the upper limit presented in Section~\ref{ssec_powspec} to constrain the product $A_{\textrm{CO}}f_{\textrm{duty}}^{1/2}$. This constraint is shown in Figure~\ref{fig_mminvsa} as a function of $M_{\textrm{min}}$, compared to several estimates from theoretical models.

\subsection{Constraints on Theoretical Models}\label{ssec_comodels}
As shown in Figure~\ref{fig_mminvsa}, our constraint on $A_{\textrm{CO}}$ falls below estimates from \cite{Lidz2011}, though those estimates are tailored for $z\gtrsim6$. Our measurement is not sensitive enough to place constraints on model A from \citet{Pullen2013} (hereafter \citetalias{Pullen2013}), or on estimates from \cite{Visbal2010} or \cite{Li2015} (not shown in the figure, though comparable to model A). Our sensitivity limit is not far above model estimates by \cite{Righi2008}; our measurement places constraints on optimistic versions of the Righi model, ruling out variants with an estimated power spectrum an order of magnitude stronger than the baseline prediction. 

We consider the models of \citetalias{Pullen2013} in a bit more detail, as their model B is closest to our upper limit. \citetalias{Pullen2013} present two models (A and B) that predict the CO power spectrum based on locally observed correlations between star formation rates, far-infrared luminosity and CO luminosity \citep{Kennicutt1998,Wang2011}. Model A utilizes a predicted relationship between halo mass and SFR for power spectrum estimates, while model B uses SFR functions based on UV and IR observations \citep{Smit2012}.

Adopting the values for $f_{\textrm{duty}}$ and $M_{\textrm{min}}$ found in \citetalias{Pullen2013} ($M_{\textrm{min}}=10^{9}M_{\odot}$; $f_{\textrm{duty}}= t_{s}/t_{\textrm{age}}(z)$, where $t_{s}$ is the star formation timescale, of order $10^8$ yr, and $t_{\textrm{age}}(z)$ is the Hubble time at a given redshift), we constrain $A_{\textrm{CO}}<1.2\times10^{-5}\ L_{\odot}\ M_{\odot}^{-1}$. Using Equation~\ref{eqn_brighttemptoA}, we translate this to a constraint on the mean brightness temperature of the CO(1-0) transition at $z\sim3$,  $\langle T_{\textrm{CO}} \rangle < 4.8\ \mu\textrm{K}$. As shown in Figure~\ref{fig_finpowspec}, our result is inconsistent (to $2\sigma$ significance) with the baseline expectation of model B from \citetalias{Pullen2013} at our average redshift of $z=2.8$

The CO abundance may change significantly over the redshift range of our measurement, owing in part to increasing metallicity and dust masses within galaxies \citep{Valiante2009}, rising feedback and quenching processes \citep{Keres2009,Wheeler2014} and the rapid depletion of neutral gas for star formation \citep{Bauermeister2010} over cosmic time. Model B of \citetalias{Pullen2013} effectively makes such a prediction, with $A_{\textrm{CO}}$ rising with decreasing redshift as one approaches the peak of cosmic star formation (as reflected by the shaded area in Figure~\ref{fig_mminvsa}). As previously discussed in Section~\ref{ssec_powspec}, we see no evidence for a change in CO abundance as a function of redshift (evaluated over $\Delta z$  of 0.12, 0.24 and 0.49). Evaluating the \citetalias{Pullen2013} models over shorter redshift ranges, we are able to rule out the baseline version of Model B to $2.1\sigma$ significance for $z=2.3{-}2.8$, and to $1.2\sigma$ significance for $z=2.8{-}3.3$.

We note that values of $f_{\textrm{duty}}$ and $M_{\textrm{min}}$ are not well-constrained; in particular, the value of $f_{\textrm{duty}}$ differs by more than an order of magnitude between different models. \citetalias{Pullen2013} argues $f_{\textrm{duty}}=t_{s}/t_{\textrm{age}}(z)$, which is  $\sim0.05$ for the redshift range of our measurement for their assumed $t_{s}=10^{8}\ \textrm{yr}$. This choice of $t_{s}$ is based on arguments that the dominant source of CO emission at high redshift ($z\gtrsim6$) will arise from galaxies undergoing extreme starburst events \citep{Lidz2011,Righi2008}. However, there are multiple observations indicating higher star formation duty cycles at $z\sim1{-}4$, with $f_{\textrm{duty}}$ approaching 100\% \citep{Noeske2007,Lee2009,Tacconi2013}. The \citet{Li2015} model implicitly assumes a 100\% duty cycle, though they introduce intrinsic scatter model parameters for the star formation rate to $L_{\textrm{CO}}$ and halo mass to star formation rate relations to allow for observed variations in halo activity. Following Equation~\ref{eqn_shotnoisemass}, our constraint on $A_{\textrm{CO}}$ depends on $f_{\textrm{duty}}^{-0.5}$, so increasing it to unity would drop our limit on $A_\textrm{CO}$ by a factor of 4, though the unaccounted scatter in halo properties noted by \citet{Li2015} would mimic a lower $f_{\textrm{duty}}$. We use the low value of $f_{\textrm{duty}}$ from \citetalias{Pullen2013} for our conservative upper limit on $A_{\textrm{CO}}$, and show the combination of $A_{\textrm{CO}}$ and $f_{\textrm{duty}}$ in Figure~\ref{fig_mminvsa}. Variations in values of $M_{\textrm{min}}$ have smaller effects on the model constraints, as can be seen in Figure~\ref{fig_mminvsa}. Uncertainties in the halo mass function (i.e., $dn/dM$) contribute insignificantly to the uncertainty. \citet{Tinker2008} found 1\% accuracy for their fitting function over the range of masses of interest for this problem, which translates into a 1\% uncertainty in the quantity of interest for determining $A_{\textrm{CO}}$, the second moment of $dn/dM$.
\subsection{Constraints on the Cosmic CO Luminosity and Density of $H_{2}$}\label{ssec_cosmich2}
Following the arguments leading up to Equation~\ref{eqn_brighttemptoA}, one may write a simple expression for the volume emissivity of CO(1-0), $\varepsilon_{\textrm{CO(1{-}0)}}$:
\begin{equation}\label{eqn_volumeemis}
\varepsilon_{\textrm{CO(1{-}0)}} = f_{\textrm{duty}}A_{\textrm{CO}}\frac{L_{\odot}}{M_{\odot}}\int_{M_{\textrm{min}}}^{\infty} M \frac{dn}{dM}\, dM.
\end{equation}
Adopting fiducial values from \citetalias{Pullen2013} ($M_{\textrm{min}}=10^{9}M_{\odot}$, $f_{\textrm{duty}}= t_{s}/t_{\textrm{age}}(z)$), from our constraint of $A_{\textrm{CO}}\leq 1.2\times10^{-5} \ L_{\odot}\ M_{\odot}^{-1}$, we calculate an upper limit for the CO(1-0) volume emissivity of $\varepsilon_{\textrm{CO(1{-}0)}} \leq 9 \times 10^{3}\ L_{\odot} \ h^{3}\, \textrm{Mpc}^{-3}$. Our measurement offers an improved constraint on the maximum cosmic CO(1-0) luminosity at $z\sim3$ presented by \cite{Walter2014} (who determined $\varepsilon_{\textrm{CO(1{-}0)}} \lesssim 2\times10^{4}\ L_{\odot} \ h^{3}\, \textrm{Mpc}^{-3}$), as it is an integrated measurement across the entire population range of galaxies, whereas \cite{Walter2014} exclude consideration of galaxies below their detection threshold.

Assuming a linear relationship between $L_{\textrm{CO}}$ and $M_{\textrm{H}_{2}}$, one can use the CO luminosity to molecular gas mass conversion factor, $\alpha_{\textrm{CO}}$, to determine an upper limit for the cosmic density of molecular gas. Assuming a Milky Way-like $\alpha_{\textrm{CO}}=4.3 \ M_{\odot}\ (\textrm{K}\ \textrm{km}\ \textrm{s}^{-1}\ \textrm{pc}^{-2})^{-1}$ \citep{Frerking1982,Dame2001} -- equivalent to $8.7\times10^{4}\ M_{\odot}\ L_{\odot}^{-1}$ \citep{Solomon1992} -- we find an upper limit for the global density of molecular gas of $\rho_{z\sim3}(M_{\textrm{H}_{2}})\leq 9 \times10^{8}\ M_{\odot}\ h^{3}\, \textrm{Mpc}^{-3}$,  which lies only a factor of $\sim2-3$ above theoretical predictions \citep{Obreschkow2009e,Lagos2011,Sargent2014}. Assuming $h=0.7$, our limit is equal to $\rho_{z\sim3}(M_{\textrm{H}_{2}})\leq 2.8 \times10^{8}\ M_{\odot}\ \textrm{Mpc}^{-3}$, in agreement with measurements made by \cite{Walter2014}, who calculate an upper limit of approximately $4\times10^{8}\ M_{\odot}\ \textrm{Mpc}^{-3}$. 

\subsection{Cosmic Variance and Limits of Significance}\label{ssec_cosmicvar}
We now consider the impact of cosmic variance on our measurement. In the shot-noise regime, the power measured is roughly proportional to the number density of emitters within the volume measured, $n_{\textrm{e}}$. The variance in the measured number density depends inversely on the total number of emitters detected over a given volume, $N_{\textrm{e}}=n_{e}V_{\textrm{z}}$. Therefore, provided with a halo mass function and a scaling relationship between $L_{\textrm{CO}}$ and halo mass, one can calculate an estimate for cosmic variance. With a total survey volume of $3.6\times10^{6}\ h^{-3}\, \textrm{Mpc}^{3}$, and assuming linear scaling between halo mass and $L_{\textrm{CO}}$ with $f_{\textrm{duty}}=t_{s}/t_{\textrm{age}}(z)$, we find that cosmic variance effectively induces an error of $\Delta P/P\approx 0.1$ in our measurement. However, we note that $f_{\textrm{duty}}$ has a strong effect on the impact of cosmic variance, and that higher values of $f_{\textrm{duty}}$ (suggested by some models, as discussed in Section~\ref{ssec_comodels}) will decrease the impact of cosmic variance. In the case of an upper limit for the power spectrum, such as the one established in this experiment, we find that the impact of cosmic variance is negligible and may safely be ignored.

\section{Conclusion}\label{sec_conclusion}
In this paper, we have constrained the power spectrum for CO at $z\sim3$ to $P_{\textrm{CO}}<2.6\times10^{4} \mu\textrm{K}^{2} (h^{-1}\,\textrm{Mpc})^{3}$, or $\Delta^{2}_{\textrm{CO}}(k\!=\!1\ h\,\textrm{Mpc}^{-1})<1.3\times10^{3}\ \mu\textrm{K}^{2}$, to $2\sigma$ confidence. We have used this constraint to place limits on the mean brightness temperature for CO and the global density of molecular gas. We have ruled out the \cite{Lidz2011} model for $z\sim3$, and have placed constraints on model B from \cite{Pullen2013}.

In recent years, additional observations -- focused on intensity mapping of CO(1-0) at $z\sim3$ -- have been performed with the SZA. Data from these observations will yield a significantly more sensitive measurement, and will offer improved constraints on the abundance of CO and molecular gas in the early Universe. Other compact low-resolution centimeter-wave instruments, such as the Yuan-Tseh Lee Array \citep{Ho2009}, will offer increased sensitivity and will be capable of deeply probing the CO power spectrum at $z\sim3$.
\acknowledgments
The authors would like to thank the referee for their timely and thoughtful feedback, which helped improve the quality and clarity of this paper. We also thank F. M. Fornasini and R. L. Plambeck for their valuable feedback during the preparation of this manuscript. This work was supported in part by the National Science Foundation University Radio Observatories Program, AST-1140031. We gratefully acknowledge the James S.\ McDonnell Foundation, the National Science Foundation and the University of Chicago for funding to construct the SZA.  The operation of the SZA was supported by NSF through award AST-0604982. Partial support was provided by NSF Physics Frontier Center grant PHY-0114422 to the Kavli Institute of Cosmological Physics at the University of Chicago, and by NSF grants AST-0507545 and AST-0507161 to Columbia University.
\appendix
\section{The Intensity Mapping Technique}\label{app_maptechnique}
The goal of intensity mapping is to use the two-point autocorrelation function, $\xi(\mathbf{x})$, to probe the underlying population of galaxies bearing molecular gas (through the tracer CO molecule) within some representative volume, $\textrm{V}_{z}$, of the Universe at redshift $z$. The final analysis product is a power spectrum, $P(k)$, which along with its dimensionless counterpart, $\Delta^{2}(k)$, is related to the two-point autocorrelation function by
\begin{eqnarray}
\Delta^{2}(k) &\equiv& \frac{k^3}{2\pi^2}P(k), \nonumber \\
&=& \frac{k^3}{2\pi^2} \left \langle \mathcal{F}_{x \rightarrow k}\left ( \xi(\mathbf{x}) \right) \right \rangle_{\mathbf{k}\cdot\mathbf{k}=k^{2}}, \label{eqn_deltafun}
\end{eqnarray}
where $\mathcal{F}_{x \rightarrow k}$ is the Fourier transform from configuration to Fourier space, and $\mathbf{k}$ is the vector wavenumber (of magnitude $k$) in Fourier space. The two point autocorrelation function is further defined as: 
\begin{eqnarray}
\xi(\mathbf{x}) &=& \frac{1}{W(\mathbf{x}) \cdot W(\mathbf{x})} \left(\frac{c^{2}}{2k_{B}\nu^{2}}\right)^{2} \left ( (W(\mathbf{x}) I(\mathbf{x^\prime})) \star (W(\mathbf{x}) I(\mathbf{x})) \right) \label{eqn_autocorr}
\end{eqnarray}
where $I(\mathbf{x})$ is the specific intensity observed at a given frequency, $\nu$. $W(\mathbf{x})$ is the 
spatial windowing function applied over the volume in question. $W(\mathbf{x})\cdot W(\mathbf{x})$ is the normalization term of the autocorrelation function, which is equal to the effective volume probed, $\textrm{V}_{z}$ . The Fourier dual of the autocorrelation function is $\tilde{\xi}(\mathbf{k})$, such that $P = \langle \tilde{\xi}(\mathbf{k}) \rangle$. The function $\tilde{\xi}(\mathbf{k})$ can then be expressed as
\begin{equation}\label{eqn_deltacomb}
\tilde{\xi}(\mathbf{k}) = \frac{1}{\textrm{V}_{z}} \left(\frac{c^{2}}{2k_{B}\nu^{2}}\right)^{2}\left( ( \tilde{W}^{*}(\mathbf{k})*\tilde{I}^{*}(\mathbf{k}) )( \tilde{W}(\mathbf{k})*\tilde{I}(\mathbf{k}))\right ),
\end{equation}
where $\tilde{I}(\mathbf{k})$ is the Fourier dual of $\tilde{I}(\mathbf{x})$. We map between comoving physical size, $\mathbf{r} = (r_x,r_y,r_z)$, and the native observing units of $(l,m,\Delta\nu)$, where $l$ and $m$ describe the angular position (in units of radians) and $\Delta\nu$ is the change in line frequency due to expansion of the Universe, with the following expressions:
\begin{equation}\label{eqn_xyztolmdnu}
l =\frac{r_x}{D_{M}(z)},\quad m=\frac{r_y}{D_{M}(z)},\quad \Delta\nu =\frac{H_{0}E(z)\nu_{rest}}{c(1+z)^2}r_z.
\end{equation}
In Equation~\ref{eqn_xyztolmdnu}, $D_{M}(z)$ is the comoving radial distance for redshift $z$, $\nu_{o}$ is the rest frequency of the line transition, $H_{0}$ is the current Hubble parameter and $E(z)$ is the dimensionless Hubble parameter. To evaluate $\tilde{\xi(\mathrm{k})}$, one converts $(u,v,\eta)$, the Fourier dual
of $(l,m,\Delta\nu)$, to $(k_{x},k_{y},k_{z})$:
\begin{align*}
u = \frac{k_{x}X}{2\pi},\quad v =\frac{k_{y}X}{2\pi},\quad \eta =\frac{k_{z}Y}{2\pi},
\end{align*}
using
\begin{equation}\label{eqn_xyconv}
X=D_{M}(z),\quad\quad Y=\frac{c(1+z)^2}{H_{0}E(z)\nu_{rest}}.
\end{equation}
In Equation~\ref{eqn_xyconv}, we have included the variables $X$ and $Y$ (defined in \citealt{Parsons2012}) for brevity only, with units of $\textrm{Mpc}\ \textrm{rad}^{-1}$ and $\textrm{Mpc}\ \textrm{Hz}^{-1}$ respectively. Rewriting Equation~\ref{eqn_deltacomb} and making the proper variable substitutions, we derive the following expression:
\begin{equation}\label{eqn_2ptfourier}
\tilde{\xi}(\mathbf{k}) = 
\frac{X^{4}Y^{2}}{\textrm{V}_{z}} \left(\frac{c^{2}}{2k_{B}\nu^{2}}\right)^{2}\left( ( \tilde{W}^{*}(u,v,\eta)*\tilde{I}^{*}(u,v,\eta) )( \tilde{W}(u,v,\eta)*\tilde{I}(u,v,\eta))\right ).
\end{equation}

\begin{figure*}[t]
\begin{center}
\includegraphics[scale=0.5]{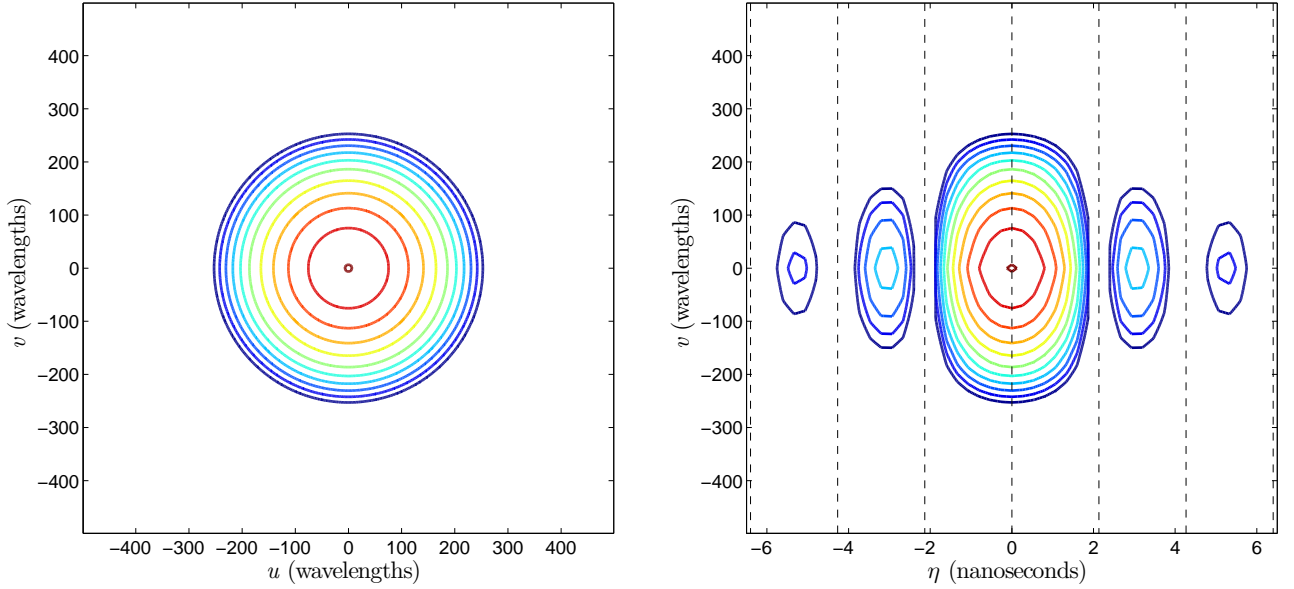}
\caption{\textit{Left}: A single cut through the function $\tilde{W}^{2}$, where $\eta=0$. Different contours correspond to 0.2 dex differences in sensitivity, from unity (dark red) to 0.01 (dark blue). \textit{Right}: Another slice through $\tilde{W}^{2}$, where $u=0$. Dashed lines mark the discrete values of $\eta$ sampled by our analysis.
\label{fig_winfun}}
\end{center}
\end{figure*}

With our simplified expression, we now turn our attention to the interferometer response to determine its relation to $\tilde{I}(u,v,\eta)$, and thus the measurement of $\tilde{\xi}(\mathbf{k})$. Under the flat-sky approximation, the interferometer response is given by
\begin{eqnarray}\label{eqn_vis}
\mathcal{V}(u,v,\nu) &=& \mathcal{F}_{(l,m)\rightarrow(u,v)}\left (I(l,m,\nu)\cdot A_{\nu}(l,m) \right ) \nonumber \\
&=&\tilde{I}(u,v,\nu)*\tilde{A}_{\nu}(u,v),
\end{eqnarray}
where measurements for the interferometer response, $\mathcal{V}(u,v,\nu)$, are commonly referred to as ``visibilities'' \citep{TMS1986}. $\mathcal{F}$ in Equation~\ref{eqn_vis} is the Fourier transform operator, which in this case transforms between $(l,m)$ and $(u,v)$ domains, and $A_{\nu}(l,m)$ is the primary beam response of the telescope at a given frequency $\nu$ and position in the sky $(l,m)$.  We wish however to move from $(u,v,\nu)$ to $(u,v,\eta)$ and create what we will refer to as ``delay-visibilities''. This requires performing one last Fourier transform over some window in frequency $\mathcal{B}(\nu)$;
\begin{eqnarray}
\tilde{\mathcal{V}}(u,v,\eta) &=& \mathcal{F}_{\nu\rightarrow\eta}\left (V(u,v,\nu)\cdot \mathcal{B}(\nu) \right ), \nonumber \\
&=& \left(\tilde{I}(u,v,\eta)*A_{\nu}(u,v)\right)*\tilde{\mathcal{B}}(\eta) =\tilde{I}(u,v,\eta)*\tilde{W}(u,v,\eta). \label{eqn_delayvis}
\end{eqnarray}
The delay-visibility in Equation~\ref{eqn_delayvis} is the convolution of $\tilde{I}$ and our windowing function, which we have defined as $\tilde{W}(u,v,\eta)=\tilde{A}(u,v)*\tilde{\mathcal{B}}(\eta)$, based on the windowing functions created by the primary beam pattern and the bandwidth window.  We define $\Omega_{\textrm{B},\nu}=\int A_{\nu}\, d\Omega$ to be the solid angle of the primary beam of the telescope, where $A_{\nu}$ is the primary beam pattern (and the Fourier dual of the aperture function, $\tilde{A}$), for a given frequency $\nu$. Nominally both $A$ and $\Omega$ are frequency dependent, but the frequency range of each window is small enough that both $\Omega_{\textrm{B},\nu}$ and $A_{\nu}$ can be treated as constant over the volume of interest. We also define $B_{z}=\int \mathcal{B}(\nu)\,d\nu$ as the effective bandwidth for our observation of the volume of interest, where $\mathcal{B}(\nu)$ is the frequency windowing function. In practice, this windowing function is generally equal to unity over the correlator window, hence for our measurement $B_{z}$ is equal to the correlator window bandwidth. We will assume that our range in redshift over which our volume $\textrm{V}_{z}$ subsides is small enough that $X$ and $Y$ remain constant. With this assumption we can further simplify our expression by substituting $\textrm{V}_{z}$ with defined instrument parameters. The effective volume can be defined as 
\begin{equation}\label{eqn_effvolume}
\textrm{V}= \int W \cdot W = X^{2}Y \int A\mathcal{B}\cdot A\mathcal{B}\, d\Omega\, d\nu = X^{2}YB_{z}\Omega_{\textrm{B},\nu}/2.
\end{equation}
In Equation~\ref{eqn_effvolume}, we have made use of the fact that for a Gaussian beam, $\int A^{2} = \Omega_{\textrm{B}}/2$, and have assumed a uniform weighting in frequency (i.e., $\mathcal{B}=1$ over the frequency range of the window, otherwise $\mathcal{B}=0$), as it is the weighting scheme used in our analysis. We can now simplify Equation~\ref{eqn_2ptfourier}, and come up with an expression for Equation~\ref{eqn_deltafun} that contains only observables:
\begin{eqnarray}
P(k) = \frac{2X^{2}Y}{\Omega_{\textrm{B}}B_{z}}\left(\frac{c^{2}}{2k_{B}\nu^{2}}\right)^{2} \left \langle \tilde{\mathcal{V}}^{*}(u,v,\eta)\tilde{\mathcal{V}}(u,v,\eta) \right \rangle_{\mathbf{k}\cdot\mathbf{k}=k^{2}} \label{eqn_powspecfinal},\\
\Delta^{2}(k) = \frac{k^{3}}{\pi^{2}} \frac{X^{2}Y}{\Omega_{\textrm{B}}B_{z}}\left(\frac{c^{2}}{2k_{B}\nu^{2}}\right)^{2} \left \langle \tilde{\mathcal{V}}^{*}(u,v,\eta)\tilde{\mathcal{V}}(u,v,\eta) \right \rangle_{\mathbf{k}\cdot\mathbf{k}=k^{2}}. \label{eqn_deltaspecfinal}
\end{eqnarray}
A subtle feature of Equations~\ref{eqn_powspecfinal} and~\ref{eqn_deltaspecfinal} is that each delay-visibility is a weighted mean of several values of $\tilde{I}^{2}(u,v,\eta)$, smeared together by the windowing function $\tilde{W}^{2}$ (shown in Figure \ref{fig_winfun}). This smearing effect in $(u,v,\eta)$ has three significant consequences worth noting.

The first consequence is that the number of independent $k$-modes of the power spectrum, $N_k$, will be limited by the size of the windowing function. The size of the `footprint' of a measurement is $\Delta\tilde{\textrm{V}} = \int W^{2}$. While this value does not directly affect the estimators $\hat{P}$ and $\hat{\Delta}^{2}$, it will affect the statistical significance of our measurement by limiting the number of independent modes one can measure. 

The second consequence is that $|\tilde{I}^{2}|$ may change appreciably over the range in $(u,v,\eta)$ that our delay-visibility is spread over, particularly in cases where the delay-visibility probes a large range in spatial scales (a scenario not applicable to the \citetalias{Sharp2010} data set, where $\Delta k < k$). If unaccounted for, this may bias estimates of the power spectrum and lead to misinterpretation of results.

The final consequence is that delay-visibilities centered at different positions in $(u,v,\eta)$ may measure partially overlapping regions of $\tilde{I}(u,v,\eta)$, and can produce a measurement of the power spectrum. Invoking the assumption that the Universe is homogeneous and isotropic on sufficiently large scales, orthogonal spatial modes for a randomized brightness temperature fluctuation field will be incoherent with one another, such that $\langle \tilde{I}^{*}(k)\tilde{I}(k^\prime) \rangle_{k \ne k^{\prime}}=0 $. The product of two delay-visibilities offset by some distance in $(u,v,\eta)$ produces a coherent measurement of the specific intensity variance, $I^{2}$, weighted by the inner product of the two offset windowing functions:
\begin{equation}\label{eqn_coherentfrac}
C(\Delta u,\Delta v, \Delta \nu) = \frac{ \int W(u,v,\eta)\cdot W(u+\Delta u,v+\Delta v,\eta + \Delta \eta) \,du\,dv\,d\eta}{\int W(u,v,\eta)\cdot W(u,v,\eta)\,du\, dv\, d\eta}.
\end{equation}
$C(\Delta u, \Delta v, \Delta \eta)$ is the normalized covariance (with respect to the astronomical signal being measured) of two delay-visibilities separated by some distance $(\Delta u, \Delta v, \Delta \eta)$, normalized such that for two delay-visibilities with zero separation have a normalized covariance of $C(0,0,0)=1$. With this we can now define our estimator, $|\hat{\tilde{I}}^{2}|$, for the true specific intensity variance, $|\tilde{I}^{2}|$, as
\begin{eqnarray}\label{eqn_specintsq}
|\hat{\tilde{I}}^{2}(u,v,\eta)| = \frac{\sum\limits_{n}^{n \ne m} \mathcal{\tilde{V}}_{m}^{*}(u,v,\eta)\mathcal{\tilde{V}}_{n}(u^\prime,v^\prime,\eta^\prime)C(u - u^\prime, v - v^\prime , \eta - \eta^\prime)\left (\sigma^2_{m}\sigma^2_{n} \right )^{-2}}
{\sum\limits_{n}^{n \ne m} \left ( C(u - u^\prime, v - v^\prime , \eta - \eta^\prime)/\sigma^2_{m}\sigma^2_{n} \right )^{2}}
\end{eqnarray}
where $\sigma^2_{m}$ is the noise estimate in a given visibility, $\mathcal{V}_{m}$. The sum is carried out for each visibility $\mathcal{V}_m (u,v,\eta)$ over all other independent visibilities $\mathcal{V}_n(u^\prime,v^\prime,\eta^\prime)$. We discard any terms that arise from the product of a delay-visibility with its own complex conjugate in order to prevent our estimator from being positively biased by instrumental noise. Our sum in Equation~\ref{eqn_specintsq} has been naturally weighted so as to maximize the sensitivity of our measurement (though other weighting schemes can be used). With this, we define our estimators for the power spectrum, $\hat{P}$ and $\hat{\Delta}^{2}$, as
\begin{eqnarray}
\hat{P}(k) = \frac{2X^{2}Y}{\Omega_{\textrm{B}}B_{z}}\left(\frac{c^{2}}{2k_{B}\nu^{2}}\right)^{2} \left \langle \left | \hat{\tilde{I}}^{2}\right | \right \rangle_{\mathbf{k}\cdot\mathbf{k}=k^{2}} \label{eqn_powspecestfinal},\\
\hat{\Delta}^{2}(k) = \frac{k^{3}}{\pi^{2}} \frac{X^{2}Y}{\Omega_{\textrm{B}}B_{z}}\left(\frac{c^{2}}{2k_{B}\nu^{2}}\right)^{2} \left \langle \left | \hat{\tilde{I}}^{2}\right | \right \rangle_{\mathbf{k}\cdot\mathbf{k}=k^{2}}. \label{eqn_deltaspecestfinal}
\end{eqnarray}
\section{Sensitivity Estimates for Intensity Mapping Experiments}\label{app_szaapprop}
Utilizing equations~\ref{eqn_powspecfinal} and~\ref{eqn_deltaspecfinal}, one may construct a simple estimate for the sensitivity one may achieve for a given choice of instruments and observations. The estimated noise for the power spectrum may be defined as
\begin{equation} \label{eqn_powspecnoiseest}
P_{N} = \frac{2X^{2}Y}{\Omega_{\textrm{B}}B_{z}}\left(\frac{c^{2}}{2k_{B}\nu^{2}}\right)^{2} \frac{\mathcal{\tilde{V}}_{\textrm{N}}^{2}}{N_{\textrm{V}}^{1/2}N_{k}^{1/2}N_{\textrm{rd}}},
\end{equation}
where $\tilde{\mathcal{V}}_{\textrm{N}}$ is the noise within a single delay-visibility, $N_{\textrm{V}}$ is the number of independent volumes measured, $N_{k}$ is the number of independent measurements made in the k-space volume, and $N_{\textrm{rd}}$ is the number of redundant measurements made at a given position of $(u,v,\eta)$. Note that $N_{\textrm{meas}} = N_{\textrm{V}}N_{\textrm{k}}$.  $\tilde{\mathcal{V}}_\textrm{N}$ can be further define as
\begin{equation}
\tilde{\mathcal{V}}_{\textrm{N}}= \frac{2k_{B}\nu^{2}}{c^{2}}\frac{\Omega_{\textrm{b}} T_{\textrm{sys}}}{\eta_{\textrm{eff}}} \sqrt{\frac{B}{\tau_{\textrm{int}}}},
\end{equation} 
where $T_{\textrm{sys}}$ is the system temperature, $\eta_{\textrm{eff}}$ is the aperture efficiency, $B$ is the bandwidth over which the measurement is being made and $\tau$ is the integration time \citep{Parsons2012,TMS1986}.

The number of independent volumes probed is simply the product of the number of fields, $N_{\textrm{f}}$, and the number of redshift windows, $N_{z}$. The number of redundant measurements, $N_\textrm{rd}$ and number of independent measurements in the k-space volume, $N_{k}$, are heavily intertwined and influenced by both the array and observing configuration. Consider an array with $N_{\textrm{base}}$ number of baselines. In a minimally redundant array, each of these baselines produce an independent measurement. However, in a redundant array (where elements are placed equally spaced grid) or in a extremely compact array, some baselines will overlap in the $uv$ plane. If we define $N_{\textrm{ob}}$ as the typical number of baselines that instantaneously share the same position in the $uv$ plane (where $N_{\textrm{ob}}=1$ corresponds to a minimally redundant array), then the number of independent measurements will decrease as $N_{\textrm{ob}}^{-1}$, but the number of redundant measurements will rise as $N_{\textrm{ob}}$. While compact, the SZA is laid out in such a way that the instantaneous baseline coverage produces only a few, partially overlapping baselines; hence, $N_{\textrm{ob,SZA}} \approx 1$. 

When the array is not coplanar with the sky, the projected baselines will change position over time -- an effect commonly referred to as Earth rotation aperture synthesis (ERAS) \citep{TMS1986}. ERAS reduces the amount of time a given baseline is found at a particular position in the $uv$ plane. Defining $f_{\textrm{r}}$ as the fraction of the integration time baselines are found at already sampled positions in the uv-plane (either because a baseline remains in a single position for that time, or the baseline moves into a new position in the $uv$ plane previously sampled by another baseline), then the number of redundant measurements scales linearly with $f_{\textrm{r}}$, though the number of independent measurements will scale as $f_{r}^{-1}$. The calculation of $f_{\textrm{r}}$ will be heavily affected by the position of the source(s) of interest, as well as the range in hour angle over which observations are conducted, but a simple estimate for redundant fraction is $f_{\textrm{r}}=2d_{\textrm{a}}^{2}/d_\textrm{B,med}^{2}$, where $d_{\textrm{A}}$ is the diameter of the antenna and $d_{\textrm{B,med}}$ is the median baseline length. The justification for this estimate is that the interferometer will sweep out an area roughly equal to $\pi d_{\textrm{B,med}}^{2}$ in the $uv$ plane, whereas the area instantaneously sampled by a single baseline is $2\pi d_{\textrm{a}}^{2}$ (the factor of 2 arises from the Hermitian nature of the $uv$ plane). Therefore, the interferometer measures approximately $d_\textrm{B,med}^{2}/2d_{\textrm{a}}^{2}$ independent positions. Excluding baselines with the outrigger antennas, the median SZA baseline is approximately 6.5 meters, leading to $f_{\textrm{r,SZA}} \approx 0.6$.

Finally, if one has $N_{\textrm{ch}}$ frequency channels across a particular redshift window, then there are an equal number of delay-channels, each of which provide an independent measurement. One caveat to this statement is that if the frequency resolution is too fine, then the flux from an individual emitter may become spread across several channels, leaving some delay-visibilities to resolve out the emission and not meaningfully contribute to the measurement (at a velocity resolution of $\sim 300 \textrm{km}/\textrm{s}$, this should not be an issue for the SZA). Combining all of the above pieces of information, the estimates for the number independent and redundant measurements may be written as,\begin{eqnarray}
N_{\textrm{rd}} &\approx& N_{\textrm{ob}} f_{\textrm{r}}, \label{eqn_numrdmeas}\\
N_{k} &\approx& N_{\textrm{base}} N_{\textrm{ob}}^{-1} f_{\textrm{r}}^{-1} N_{\textrm{ch}} \label{eqn_numindmeas}.
\end{eqnarray}
Combining Equations~\ref{eqn_numrdmeas} and~\ref{eqn_numindmeas} with Equation~\ref{eqn_powspecnoiseest}:
\begin{equation}
P_{N} \sim 2X^{2}Y\Omega_{\textrm{B}}\frac{(T_{\textrm{sys}}/\eta_{\textrm{eff}})^{2}}{\tau_{\textrm{int}}(2N_{\textrm{fields}}N_{z}N_{\textrm{base}}N_{\textrm{ob}}N_{\textrm{ch}})^{1/2}}\frac{d_{\textrm{B,med}}}{d{_\textrm{a}}}.
\end{equation}
Plugging in values from the \citetalias{Sharp2010} data set, we obtain an estimated sensitivity of $P_{N}\sim 1.3\times 10^{4} \mu\textrm{K}^{2} (h^{-1}\ \textrm{Mpc})^{3}$, very close to the actual sensitivity achieved in Section~\ref{sec_results}.
\bibliography{legacybib}

\begin{thebibliography}{}
\expandafter\ifx\csname natexlab\endcsname\relax\def\natexlab#1{#1}\fi

\bibitem[{{Bauermeister} {et~al.}(2010){Bauermeister}, {Blitz}, \&
  {Ma}}]{Bauermeister2010}
{Bauermeister}, A., {Blitz}, L., \& {Ma}, C.-P. 2010, \apj, 717, 323

\bibitem[{{Bennett} {et~al.}(2013){Bennett}, {Larson}, {Weiland}, {Jarosik},
  {Hinshaw}, {Odegard}, {Smith}, {Hill}, {Gold}, {Halpern}, {Komatsu}, {Nolta},
  {Page}, {Spergel}, {Wollack}, {Dunkley}, {Kogut}, {Limon}, {Meyer}, {Tucker},
  \& {Wright}}]{Bennett2013}
{Bennett}, C.~L., {Larson}, D., {Weiland}, J.~L., {et~al.} 2013, \apjs, 208, 20

\bibitem[{{Bolatto} {et~al.}(2013){Bolatto}, {Wolfire}, \&
  {Leroy}}]{Bolatto2013}
{Bolatto}, A.~D., {Wolfire}, M., \& {Leroy}, A.~K. 2013, \araa, 51, 207

\bibitem[{{Bouwens} {et~al.}(2012){Bouwens}, {Illingworth}, {Oesch}, {Trenti},
  {Labb{\'e}}, {Franx}, {Stiavelli}, {Carollo}, {van Dokkum}, \&
  {Magee}}]{Bouwens2012}
{Bouwens}, R.~J., {Illingworth}, G.~D., {Oesch}, P.~A., {et~al.} 2012, \apjl,
  752, L5

\bibitem[{{Breysse} {et~al.}(2014){Breysse}, {Kovetz}, \&
  {Kamionkowski}}]{Breysse2014}
{Breysse}, P.~C., {Kovetz}, E.~D., \& {Kamionkowski}, M. 2014, \mnras, 443,
  3506

\bibitem[{{Carilli}(2011)}]{Carilli2011}
{Carilli}, C.~L. 2011, \apjl, 730, L30

\bibitem[{{Dame} {et~al.}(2001){Dame}, {Hartmann}, \& {Thaddeus}}]{Dame2001}
{Dame}, T.~M., {Hartmann}, D., \& {Thaddeus}, P. 2001, \apj, 547, 792

\bibitem[{{Datta} {et~al.}(2010){Datta}, {Bowman}, \& {Carilli}}]{Datta2010}
{Datta}, A., {Bowman}, J.~D., \& {Carilli}, C.~L. 2010, \apj, 724, 526

\bibitem[{{Decarli} {et~al.}(2014){Decarli}, {Walter}, {Carilli}, {Riechers},
  {Cox}, {Neri}, {Aravena}, {Bell}, {Bertoldi}, {Colombo}, {Da Cunha}, {Daddi},
  {Dickinson}, {Downes}, {Ellis}, {Lentati}, {Maiolino}, {Menten}, {Rix},
  {Sargent}, {Stark}, {Weiner}, \& {Weiss}}]{Decarli2014}
{Decarli}, R., {Walter}, F., {Carilli}, C., {et~al.} 2014, \apj, 782, 78

\bibitem[{{Dickinson} {et~al.}(2003){Dickinson}, {Giavalisco}, \& {GOODS
  Team}}]{Dickinson2003}
{Dickinson}, M., {Giavalisco}, M., \& {GOODS Team}. 2003, in The Mass of
  Galaxies at Low and High Redshift, ed. R.~{Bender} \& A.~{Renzini}, 324

\bibitem[{{Dodelson}(2003)}]{Dodelson2003}
{Dodelson}, S. 2003, {Modern cosmology}

\bibitem[{{Downes} {et~al.}(1993){Downes}, {Solomon}, \&
  {Radford}}]{Downes1993}
{Downes}, D., {Solomon}, P.~M., \& {Radford}, S.~J.~E. 1993, \apjl, 414, L13

\bibitem[{{Frerking} {et~al.}(1982){Frerking}, {Langer}, \&
  {Wilson}}]{Frerking1982}
{Frerking}, M.~A., {Langer}, W.~D., \& {Wilson}, R.~W. 1982, \apj, 262, 590

\bibitem[{{Furlanetto} {et~al.}(2006){Furlanetto}, {Oh}, \&
  {Briggs}}]{Furlanetto2006}
{Furlanetto}, S.~R., {Oh}, S.~P., \& {Briggs}, F.~H. 2006, \physrep, 433, 181

\bibitem[{{Genzel} {et~al.}(2012){Genzel}, {Tacconi}, {Combes}, {Bolatto},
  {Neri}, {Sternberg}, {Cooper}, {Bouch{\'e}}, {Bournaud}, {Burkert},
  {Comerford}, {Cox}, {Davis}, {F{\"o}rster Schreiber}, {Garcia-Burillo},
  {Gracia-Carpio}, {Lutz}, {Naab}, {Newman}, {Saintonge}, {Shapiro}, {Shapley},
  \& {Weiner}}]{Genzel2012}
{Genzel}, R., {Tacconi}, L.~J., {Combes}, F., {et~al.} 2012, \apj, 746, 69

\bibitem[{{Glover} \& {Mac Low}(2011)}]{Glover2011}
{Glover}, S.~C.~O., \& {Mac Low}, M.-M. 2011, \mnras, 412, 337

\bibitem[{{Gong} {et~al.}(2011){Gong}, {Cooray}, {Silva}, {Santos}, \&
  {Lubin}}]{Gong2011}
{Gong}, Y., {Cooray}, A., {Silva}, M.~B., {Santos}, M.~G., \& {Lubin}, P. 2011,
  \apjl, 728, L46

\bibitem[{{Ho} {et~al.}(2009){Ho}, {Altamirano}, {Chang}, {Chang}, {Chang},
  {Chen}, {Chen}, {Chen}, {Han}, {Ho}, {Huang}, {Hwang}, {Iba{\~n}ez-Romano},
  {Jiang}, {Koch}, {Kubo}, {Li}, {Lim}, {Lin}, {Liu}, {Lo}, {Ma}, {Martin},
  {Martin-Cocher}, {Molnar}, {Ng}, {Nishioka}, {O'Connell}, {Oshiro}, {Patt},
  {Raffin}, {Umetsu}, {Wei}, {Wu}, {Chiueh}, {Chiueh}, {Chu}, {Huang}, {Hwang},
  {Liao}, {Lien}, {Wang}, {Wang}, {Wei}, {Yang}, {Kesteven}, {Kingsley},
  {Sinclair}, {Wilson}, {Birkinshaw}, {Liang}, {Lancaster}, {Park}, {Pen}, \&
  {Peterson}}]{Ho2009}
{Ho}, P.~T.~P., {Altamirano}, P., {Chang}, C.-H., {et~al.} 2009, \apj, 694,
  1610

\bibitem[{{H{\"o}gbom}(1974)}]{Hogbom1974}
{H{\"o}gbom}, J.~A. 1974, \aaps, 15, 417

\bibitem[{{Hopkins} \& {Beacom}(2006)}]{Hopkins2006}
{Hopkins}, A.~M., \& {Beacom}, J.~F. 2006, \apj, 651, 142

\bibitem[{{Israel}(1997)}]{Israel1997}
{Israel}, F.~P. 1997, \aap, 328, 471

\bibitem[{{Kennicutt}(1998)}]{Kennicutt1998}
{Kennicutt}, Jr., R.~C. 1998, \apj, 498, 541

\bibitem[{{Kere{\v s}} {et~al.}(2009){Kere{\v s}}, {Katz}, {Fardal},
  {Dav{\'e}}, \& {Weinberg}}]{Keres2009}
{Kere{\v s}}, D., {Katz}, N., {Fardal}, M., {Dav{\'e}}, R., \& {Weinberg},
  D.~H. 2009, \mnras, 395, 160

\bibitem[{{Lagos} {et~al.}(2011){Lagos}, {Baugh}, {Lacey}, {Benson}, {Kim}, \&
  {Power}}]{Lagos2011}
{Lagos}, C.~D.~P., {Baugh}, C.~M., {Lacey}, C.~G., {et~al.} 2011, \mnras, 418,
  1649

\bibitem[{{Lee} {et~al.}(2009){Lee}, {Giavalisco}, {Conroy}, {Wechsler},
  {Ferguson}, {Somerville}, {Dickinson}, \& {Urry}}]{Lee2009}
{Lee}, K.-S., {Giavalisco}, M., {Conroy}, C., {et~al.} 2009, \apj, 695, 368

\bibitem[{{Li} {et~al.}(2015){Li}, {Wechsler}, {Devaraj}, \& {Church}}]{Li2015}
{Li}, T.~Y., {Wechsler}, R.~H., {Devaraj}, K., \& {Church}, S.~E. 2015, ArXiv
  e-prints, arXiv:1503.08833

\bibitem[{{Lidz} {et~al.}(2011){Lidz}, {Furlanetto}, {Oh}, {Aguirre}, {Chang},
  {Dor{\'e}}, \& {Pritchard}}]{Lidz2011}
{Lidz}, A., {Furlanetto}, S.~R., {Oh}, S.~P., {et~al.} 2011, \apj, 741, 70

\bibitem[{{Morales} {et~al.}(2012){Morales}, {Hazelton}, {Sullivan}, \&
  {Beardsley}}]{Morales2012}
{Morales}, M.~F., {Hazelton}, B., {Sullivan}, I., \& {Beardsley}, A. 2012,
  \apj, 752, 137

\bibitem[{{Mu{\~n}oz} \& {Furlanetto}(2014)}]{Munoz2014}
{Mu{\~n}oz}, J.~A., \& {Furlanetto}, S.~R. 2014, \mnras, 438, 2483

\bibitem[{{Muchovej} {et~al.}(2007){Muchovej}, {Mroczkowski}, {Carlstrom},
  {Cartwright}, {Greer}, {Hennessy}, {Loh}, {Pryke}, {Reddall}, {Runyan},
  {Sharp}, {Hawkins}, {Lamb}, {Woody}, {Joy}, {Leitch}, \&
  {Miller}}]{Muchovej2007}
{Muchovej}, S., {Mroczkowski}, T., {Carlstrom}, J.~E., {et~al.} 2007, \apj,
  663, 708

\bibitem[{{Muchovej} {et~al.}(2010){Muchovej}, {Leitch}, {Carlstrom},
  {Culverhouse}, {Greer}, {Hawkins}, {Hennessy}, {Joy}, {Lamb}, {Loh},
  {Marrone}, {Miller}, {Mroczkowski}, {Pryke}, {Sharp}, \&
  {Woody}}]{Muchovej2010}
{Muchovej}, S., {Leitch}, E., {Carlstrom}, J.~E., {et~al.} 2010, \apj, 716, 521

\bibitem[{{Noeske} {et~al.}(2007){Noeske}, {Weiner}, {Faber}, {Papovich},
  {Koo}, {Somerville}, {Bundy}, {Conselice}, {Newman}, {Schiminovich}, {Le
  Floc'h}, {Coil}, {Rieke}, {Lotz}, {Primack}, {Barmby}, {Cooper}, {Davis},
  {Ellis}, {Fazio}, {Guhathakurta}, {Huang}, {Kassin}, {Martin}, {Phillips},
  {Rich}, {Small}, {Willmer}, \& {Wilson}}]{Noeske2007}
{Noeske}, K.~G., {Weiner}, B.~J., {Faber}, S.~M., {et~al.} 2007, \apjl, 660,
  L43

\bibitem[{{Obreschkow} {et~al.}(2009{\natexlab{a}}){Obreschkow}, {Croton}, {De
  Lucia}, {Khochfar}, \& {Rawlings}}]{Obreschkow2009e}
{Obreschkow}, D., {Croton}, D., {De Lucia}, G., {Khochfar}, S., \& {Rawlings},
  S. 2009{\natexlab{a}}, \apj, 698, 1467

\bibitem[{{Obreschkow} {et~al.}(2009{\natexlab{b}}){Obreschkow}, {Heywood},
  {Kl{\"o}ckner}, \& {Rawlings}}]{Obreschkow2009b}
{Obreschkow}, D., {Heywood}, I., {Kl{\"o}ckner}, H.-R., \& {Rawlings}, S.
  2009{\natexlab{b}}, \apj, 702, 1321

\bibitem[{{Parsons} {et~al.}(2012){Parsons}, {Pober}, {McQuinn}, {Jacobs}, \&
  {Aguirre}}]{Parsons2012}
{Parsons}, A., {Pober}, J., {McQuinn}, M., {Jacobs}, D., \& {Aguirre}, J. 2012,
  \apj, 753, 81

\bibitem[{{Planck Collaboration} {et~al.}(2015){Planck Collaboration},
  {Aghanim}, {Arnaud}, {Ashdown}, {Aumont}, {Baccigalupi}, {Banday},
  {Barreiro}, {Bartlett}, {Bartolo}, \& et~al.}]{PlanckXI2015}
{Planck Collaboration}, {Aghanim}, N., {Arnaud}, M., {et~al.} 2015, ArXiv
  e-prints, arXiv:1507.02704

\bibitem[{{Pullen} {et~al.}(2013){Pullen}, {Chang}, {Dor{\'e}}, \&
  {Lidz}}]{Pullen2013}
{Pullen}, A.~R., {Chang}, T.-C., {Dor{\'e}}, O., \& {Lidz}, A. 2013, \apj, 768,
  15

\bibitem[{{Regan} {et~al.}(2001){Regan}, {Thornley}, {Helfer}, {Sheth}, {Wong},
  {Vogel}, {Blitz}, \& {Bock}}]{Regan2001}
{Regan}, M.~W., {Thornley}, M.~D., {Helfer}, T.~T., {et~al.} 2001, \apj, 561,
  218

\bibitem[{{Righi} {et~al.}(2008){Righi}, {Hern{\'a}ndez-Monteagudo}, \&
  {Sunyaev}}]{Righi2008}
{Righi}, M., {Hern{\'a}ndez-Monteagudo}, C., \& {Sunyaev}, R.~A. 2008, \aap,
  489, 489

\bibitem[{{Rudy} {et~al.}(1987){Rudy}, {Muhleman}, {Berge}, {Jakosky}, \&
  {Christensen}}]{Rudy1987}
{Rudy}, D.~J., {Muhleman}, D.~O., {Berge}, G.~L., {Jakosky}, B.~M., \&
  {Christensen}, P.~R. 1987, \icarus, 71, 159

\bibitem[{{Sargent} {et~al.}(2014){Sargent}, {Daddi}, {B{\'e}thermin},
  {Aussel}, {Magdis}, {Hwang}, {Juneau}, {Elbaz}, \& {da Cunha}}]{Sargent2014}
{Sargent}, M.~T., {Daddi}, E., {B{\'e}thermin}, M., {et~al.} 2014, \apj, 793,
  19

\bibitem[{{Schmidt}(1959)}]{Schmidt1959}
{Schmidt}, M. 1959, \apj, 129, 243

\bibitem[{{Sharp} {et~al.}(2010){Sharp}, {Marrone}, {Carlstrom}, {Culverhouse},
  {Greer}, {Hawkins}, {Hennessy}, {Joy}, {Lamb}, {Leitch}, {Loh}, {Miller},
  {Mroczkowski}, {Muchovej}, {Pryke}, \& {Woody}}]{Sharp2010}
{Sharp}, M.~K., {Marrone}, D.~P., {Carlstrom}, J.~E., {et~al.} 2010, \apj, 713,
  82

\bibitem[{{Smit} {et~al.}(2012){Smit}, {Bouwens}, {Franx}, {Illingworth},
  {Labb{\'e}}, {Oesch}, \& {van Dokkum}}]{Smit2012}
{Smit}, R., {Bouwens}, R.~J., {Franx}, M., {et~al.} 2012, \apj, 756, 14

\bibitem[{{Solomon} {et~al.}(1992){Solomon}, {Downes}, \&
  {Radford}}]{Solomon1992}
{Solomon}, P.~M., {Downes}, D., \& {Radford}, S.~J.~E. 1992, \apjl, 398, L29

\bibitem[{{Solomon} {et~al.}(1997){Solomon}, {Downes}, {Radford}, \&
  {Barrett}}]{Solomon1997}
{Solomon}, P.~M., {Downes}, D., {Radford}, S.~J.~E., \& {Barrett}, J.~W. 1997,
  \apj, 478, 144

\bibitem[{{Tacconi} {et~al.}(2013){Tacconi}, {Neri}, {Genzel}, {Combes},
  {Bolatto}, {Cooper}, {Wuyts}, {Bournaud}, {Burkert}, {Comerford}, {Cox},
  {Davis}, {F{\"o}rster Schreiber}, {Garc{\'{\i}}a-Burillo}, {Gracia-Carpio},
  {Lutz}, {Naab}, {Newman}, {Omont}, {Saintonge}, {Shapiro Griffin}, {Shapley},
  {Sternberg}, \& {Weiner}}]{Tacconi2013}
{Tacconi}, L.~J., {Neri}, R., {Genzel}, R., {et~al.} 2013, \apj, 768, 74

\bibitem[{{Thompson} {et~al.}(1986){Thompson}, {Moran}, \& {Swenson}}]{TMS1986}
{Thompson}, A.~R., {Moran}, J.~M., \& {Swenson}, G.~W. 1986, {Interferometry
  and synthesis in radio astronomy}

\bibitem[{{Tinker} {et~al.}(2008){Tinker}, {Kravtsov}, {Klypin}, {Abazajian},
  {Warren}, {Yepes}, {Gottl{\"o}ber}, \& {Holz}}]{Tinker2008}
{Tinker}, J., {Kravtsov}, A.~V., {Klypin}, A., {et~al.} 2008, \apj, 688, 709

\bibitem[{{Valiante} {et~al.}(2009){Valiante}, {Schneider}, {Bianchi}, \&
  {Andersen}}]{Valiante2009}
{Valiante}, R., {Schneider}, R., {Bianchi}, S., \& {Andersen}, A.~C. 2009,
  \mnras, 397, 1661

\bibitem[{{Visbal} \& {Loeb}(2010)}]{Visbal2010}
{Visbal}, E., \& {Loeb}, A. 2010, \jcap, 11, 16

\bibitem[{{Visbal} {et~al.}(2011){Visbal}, {Trac}, \& {Loeb}}]{Visbal2011}
{Visbal}, E., {Trac}, H., \& {Loeb}, A. 2011, \jcap, 8, 10

\bibitem[{{Walter} {et~al.}(2014){Walter}, {Decarli}, {Sargent}, {Carilli},
  {Dickinson}, {Riechers}, {Ellis}, {Stark}, {Weiner}, {Aravena}, {Bell},
  {Bertoldi}, {Cox}, {Da Cunha}, {Daddi}, {Downes}, {Lentati}, {Maiolino},
  {Menten}, {Neri}, {Rix}, \& {Weiss}}]{Walter2014}
{Walter}, F., {Decarli}, R., {Sargent}, M., {et~al.} 2014, \apj, 782, 79

\bibitem[{{Wang} {et~al.}(2011){Wang}, {Wagg}, {Carilli}, {Neri}, {Walter},
  {Omont}, {Riechers}, {Bertoldi}, {Menten}, {Cox}, {Strauss}, {Fan}, \&
  {Jiang}}]{Wang2011}
{Wang}, R., {Wagg}, J., {Carilli}, C.~L., {et~al.} 2011, \aj, 142, 101

\bibitem[{{Wheeler} {et~al.}(2014){Wheeler}, {Phillips}, {Cooper},
  {Boylan-Kolchin}, \& {Bullock}}]{Wheeler2014}
{Wheeler}, C., {Phillips}, J.~I., {Cooper}, M.~C., {Boylan-Kolchin}, M., \&
  {Bullock}, J.~S. 2014, \mnras, 442, 1396

\bibitem[{{Wilson} {et~al.}(1970){Wilson}, {Jefferts}, \&
  {Penzias}}]{Wilson1970}
{Wilson}, R.~W., {Jefferts}, K.~B., \& {Penzias}, A.~A. 1970, \apjl, 161, L43

\bibitem[{{Wolfire} {et~al.}(2010){Wolfire}, {Hollenbach}, \&
  {McKee}}]{Wolfire2010}
{Wolfire}, M.~G., {Hollenbach}, D., \& {McKee}, C.~F. 2010, \apj, 716, 1191

\bibitem[{{Young} \& {Scoville}(1982)}]{Young1982}
{Young}, J.~S., \& {Scoville}, N. 1982, \apj, 258, 467

\bibitem[{{Young} {et~al.}(1995){Young}, {Xie}, {Tacconi}, {Knezek}, {Viscuso},
  {Tacconi-Garman}, {Scoville}, {Schneider}, {Schloerb}, {Lord}, {Lesser},
  {Kenney}, {Huang}, {Devereux}, {Claussen}, {Case}, {Carpenter}, {Berry}, \&
  {Allen}}]{Young1995}
{Young}, J.~S., {Xie}, S., {Tacconi}, L., {et~al.} 1995, \apjs, 98, 219

\end{thebibliography}
\end{document}